\documentclass[
aps,
prd,
tightenlines,
groupedaddress,
showpacs,
preprintnumbers,
nofootinbib,
nobibnotes,
notitlepage,
amsmath,
amssymb, 
onecolumn,
11pt,
floatfix
]{revtex4-1}

\usepackage[utf8]{inputenc}
\usepackage[normalem]{ulem}
\usepackage{graphicx}% Include figure files
\usepackage{dcolumn}% Align table columns on decimal point
\usepackage[colorlinks=true,allcolors=purple]{hyperref}
\usepackage{url}
\usepackage{enumerate}
\usepackage{comment}

\usepackage{slashed,multirow,relsize,soul,feynmp-auto,tikz}
\usepackage{color}
\usepackage{mathrsfs} % pretty maths
\usepackage{amsmath}
\usepackage{cancel}
\usepackage{bbold}
\usepackage{mathrsfs}
\usepackage{braket}
\usepackage{physics}
\usepackage{siunitx}
\usepackage{multirow}
\usepackage[capitalize]{cleveref}
\usepackage{xspace}
\usepackage{float}

\usepackage{fontawesome} % for code link icons
\definecolor{nicegreen}{rgb}{0., 0.75, 0.46}

\definecolor{palatinate}{rgb}{0.494, 0.192, 0.482}
\definecolor{blue-violet}{rgb}{0.33, 0.17, 0.89}

\usepackage{ulem}

\begin{document}

\preprint{MI-HET-873}

\title{Long-lived particle production through the PRISM}

\author{Kevin J.~Kelly}
\email{kjkelly@tamu.edu}
\affiliation{Mitchell Institute for Fundamental Physics and Astronomy,\\
             Department of Physics and Astronomy, Texas A\&M
             University, College Station, USA}

\author{Mudit Rai}
\email{muditrai@tamu.edu}
\affiliation{Mitchell Institute for Fundamental Physics and Astronomy,\\
             Department of Physics and Astronomy, Texas A\&M
             University, College Station, USA}

\date{\today}
%==================================================================================

%==================================================================================
\begin{abstract}
Accelerator-based neutrino experiments offer a competitive environment to search for long-lived particles with sub-GeV masses. Yet, many theoretical models involving such particles predict very similar phenomenology and nearly identical final-state signatures. In view of this, we study the capabilities of upcoming experiments -- specifically the DUNE near detectors -- to distinguish between different classes of long-lived particles and the mechanisms by which they are produces. We expound how the experiment's excellent energy resolution, combined with the possibility to move the detector off-axis (the DUNE-PRISM concept), work in tandem to improve the discrimination power.
\end{abstract}
%==================================================================================

%==================================================================================
\maketitle
%==================================================================================

%==================================================================================
\section{Introduction}
\label{sec:Introduction}
%==================================================================================

The search for phenomena beyond the Standard Model (SM) of particle physics is wide-ranging, comprising searches for the mechanism behind neutrino masses, considerations of interactions between dark matter and the SM, and endeavors for precision understanding of the Higgs mechanism, among others.
Increasing the likelihood of such beyond-SM (BSM) discoveries requires casting a wide net, and fully leveraging existing and future experimental facilities to carry out as many distinct searches as possible.
Terrestrial neutrino facilities -- many sourced by high-intensity, high-energy proton beams -- offer a plethora of possible new-physics searches.
This is especially true of neutrino near-detector facilities, awash in a large flux of neutrinos and, potentially, BSM particles that can leave novel signatures in detectors.

Following the discussion of many recent studies (see, e.g., Ref.~\cite{Batell:2022dpx, Gori:2022vri} for in-depth discussion), we are interested in the prospects of discovering one or more new-physics particles in neutrino facilities in the coming decade(s). Some of the most exciting prospects for discovery come from long-lived particles (LLPs), motivated in many specific BSM contexts, where such a particle is produced associated with the high-intensity proton-target collisions\footnote{Many specific production modes exist, and we will discuss several concrete ones in this work. Other, more exotic scenarios (e.g., production of a new, long-lived state due to neutrino upscattering in the nearby matter), may be applicable as well.} and travels to the neutrino detector, decaying inside. The advantage here is that the decay signatures of such long-lived particles are typically distinct enough from the expected neutrino-scattering type of backgrounds, allowing for low-background searches to be performed simultaneously with the neutrino-beam operation of these facilities.

Many scenarios predict such LLP production in these contexts, arising from a number of distinct production mechanisms. These include (but are not limited to)
\begin{itemize}
  \item Neutral/charged meson decays,
  \item Proton bremsstrahlung,
  \item Electromagnetic interactions (e.g.\ $e^\pm$ bremsstrahlung, $e^+ e^-$ annihilation...).
\end{itemize}
If working in a specific new-physics model, then some linear combination of these production mechanisms is predicted, typically as a function of the LLP mass. For instance, if working with a ``secluded dark photon''\footnote{Here we imagine a particle with a mass $m_{A^\prime}$ that interacts with the SM purely through kinetic mixing with the SM $U(1)$ gauge group.}, the relative importance of these production mechanisms will depend on detector considerations as well as the dark photon's mass. Typically at low masses, the decay $\pi^0 \to \gamma A^\prime$ dominates, where other mechanisms become more relevant as the mass increases. 

One way of detecting the existence of these LLPs is through their decays to SM particles in the neutrino detector(s) of interest. Neutrino detectors that are situated in proton-driven neutrino beams tend to be designed to detect outgoing charged particles as well as photons. With particle identification (separation of muons and pions, or electrons and photons, for instance) as a benefit, searches for LLP decays allow for a rich set of signal topologies. As with the production mechanisms depending on the LLPs' mass(es), so do the potential decay channels. Especially when restricting to ``viable'' decay channels (i.e.\ ones that a modern neutrino detector can identify and meaningfully separate from background events), this becomes more evident. 
For instance, considering only states detectable in a neutrino detector, a LLP with a mass below $\sim\SI{100}{MeV}$ can only decay into photons and/or electrons.
This mass range is of particular importance for the upcoming neutrino-facility dark sectors program, which has relative strength against other LLP searches in the sub-GeV regime due to the beam energies and detector designs present.

Often, the specific models of interest will yield common signatures, e.g.\ the LLP decaying into an electron/positron pair in the detector -- such phenomenology is present when considering dark photons, Higgs-portal scalars, leptophilic gauge bosons, and more. If allowing for additional invisible particles (i.e.\ neutrinos) in the final state, then heavy neutral lepton models will also predict the same phenomena. With that in mind, and with hopeful discoveries of these anomalous signals on the horizon, it is wise to consider our capabilities for distinguishing between different models which predict the same experimental signatures.

A plethora of neutrino facilities are currently operating or soon will be; many of which are suitable for carrying out these searches. For specificity, we will focus our study on the upcoming Deep Underground Neutrino Experiment (DUNE) and its near detectors. This is because the DUNE program naturally aligns with many of the hallmarks of these searches -- a relatively high-energy proton beam (120 GeV), detector distances on the order of several hundred meters, and relatively large detectors with highly capable particle identification and energy/momentum reconstruction. These traits will be further bolstered if the DUNE Multi-Purpose Detector is constructed as part of the experiment's Phase II, where a magnetized gaseous-argon time-projection chamber will allow for many world-leading BSM searches. Lastly, we will also demonstrate how the planned DUNE-PRISM strategy (where the near detector moves to sample the transverse direction of the beam) will prove useful in our endeavor of distinguishing new-physics scenarios.

The remainder of this work is organized as follows. We begin with some discussion of the experimental setup and context of these searches in~\cref{sec:exp_setup}, and then in~\cref{sec:Models} we discuss the models (both in terms of UV-complete theories and in terms of phenomenologically-minded simplified frameworks) that we will consider in this work. We also present the details of the Monte Carlo simulation underpinning this study. \Cref{sec:Results} presents our analysis and potential for distinguishing among these scenarios, and~\cref{sec:Conclusions} offers some concluding remarks.

%==================================================================================
\section{Experimental setup \& Overall Motivation}
\label{sec:exp_setup}
%==================================================================================
We will explore models leading to decay signatures from long-lived particles (LLP) at the DUNE Near Detector (ND).
In addition to analyzing the signatures at the detector placed along the beam, we will also utilize the DUNE-PRISM design, which allows some of the near detectors to be moved away from the beam axis.
In the past, these signatures have been well studied in the context of specific model building, where one starts from the Lagrangian and works out the signal events and backgrounds. 
Instead of this standard approach, we intend to instead highlight common signal aspects that cut across various models.
This can be useful in identifying the particular production mechanism(s) giving rise to the LLP signature in the detector.

In order to remain concrete, we will restrict ourselves to considering renormalizable dark-sector models -- dark scalars, dark photons, etc. -- and assume that any other LLP model will be subdominant in the mass range of interest.
Since we will be analyzing DUNE sensitivity, we will consider LLPs lighter than $m_X \lesssim 1$~GeV, and heavier than the MeV scale (so that visible decays can produce particles that the DUNE ND can detect).

DUNE, primarily designed for producing an intense, collimated neutrino beam, operates in a similar fashion to many proton fixed-target experiments.
Its near-detector complex consists of a $120$~GeV proton beam striking a graphite target; proton-proton collisions then lead to production of SM mesons.
If LLPs exist, they may be produced in rare meson decays or directly from the proton beam (e.g.~from proton bremsstrahlung).
The resulting LLPs, depending on their trajectory and lifetime, may reach the DUNE ND and decay inside.
We are interested in distinct signatures which are challenging for neutrino-scattering backgrounds to mimic.
In this work, we are focused on distinguishing between different classes of new-physics production in such a facility -- we refer the interested reader to Refs.~\cite{Coloma:2023oxx,Brdar:2025hqi} for more thorough analysis of separating signal from background in this context.

The DUNE ND also benefits from the DUNE-PRISM technique, where two primary detector components can move up to 30~m in the direction transverse of the beam.
This is very useful in dark-sector identification, as different production mechanisms will lead to more-or-less geometrically focused LLP fluxes.
This comes somewhat from model specifics (e.g., the collinear enhancements present in proton-proton bremsstrahlung), and somewhat from the parent SM particle leading to the LLP production -- some SM particles, such as charged kaons, are focused in the neutrino beam whereas others are not.
Even taking data purely on-axis, the $\mathcal{O}$(few) meters radial span of the ND will allow for some differentiation due to this feature.

%==================================================================================
\section{Theoretical Framework}
\label{sec:Models}
%==================================================================================
In this section, we detail the theoretical framework and background information regarding DUNE's long-lived particle sensitivity.
We first provide a brief summary of specific benchmark models, which we will use as inspiration for the remainder of the work. 
In doing so, we will explore the parameter space in which DUNE offers novel sensitivity, motivating the mass range and lifetimes of interest for such analyses.
After doing so, we provide details of the more model-independent analysis framework that we will utilize for the remainder of this work.

%----------------------------------------------------------------------------------
\subsection{Benchmark models}
%----------------------------------------------------------------------------------
We focus on three benchmark models with qualitatively distinct LLP production mechanisms, but leading to very similar signatures in the detector.

\textbf{Dark Higgs.}  
This model introduces a light scalar field $\phi$ (the ``dark scalar'') that mixes with the Standard Model (SM) Higgs boson through a small mixing angle $\theta$ \cite{Gunion:1989we, Donoghue:1990xh, Bezrukov:2009yw, Winkler:2018qyg, Boiarska:2019jym, Batell:2020vqn}.
The phenomenology of the model is mostly characterized by a two-dimensional parameter space spanned by the scalar mass $m_\phi$ and the mixing angle $\theta$.
Due to its mixing with the Higgs boson, the dark scalar acquires couplings to SM fermions that are proportional to the corresponding SM Yukawa couplings, scaled by the mixing angle.
If $\theta$ is sufficiently small, $\phi$ decays may be suppressed to extent that the scalar becomes long-lived.

The dominant production mechanisms of $\phi$ in a proton fixed-target setup such as a neutrino beam's target station depend sensitively on the $\phi$ mass.
For $m_\phi$ below the kaon--pion mass difference, $m_K - m_\pi \simeq \SI{354}{MeV}$, production is primarily driven by two-body decays of charged mesons, in particular $K^\pm \to \pi^\pm \phi$.
If $m_\phi > m_K - m_\pi$, production in meson decays becomes kinematically forbidden, and proton bremsstrahlung processes such as $p p \to p p \phi$ become the dominant production mode. 

\begin{figure}[!htbp]
\begin{center}
\includegraphics[width=0.48\linewidth]{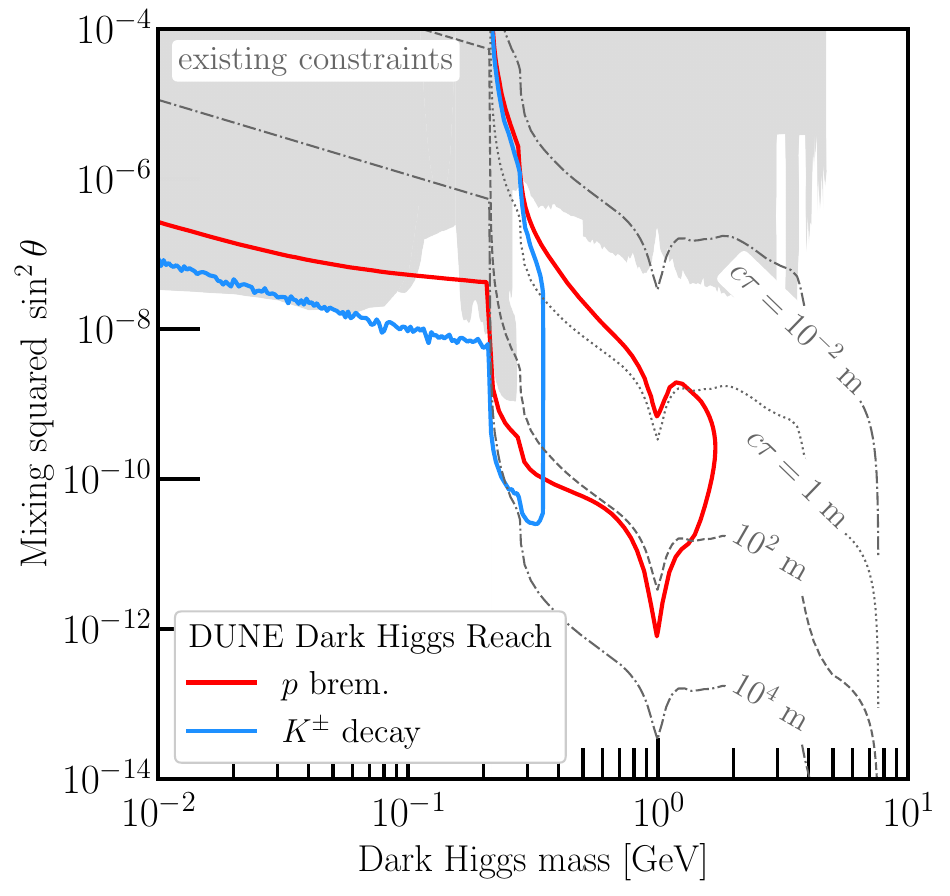}
\includegraphics[width=0.48\linewidth]{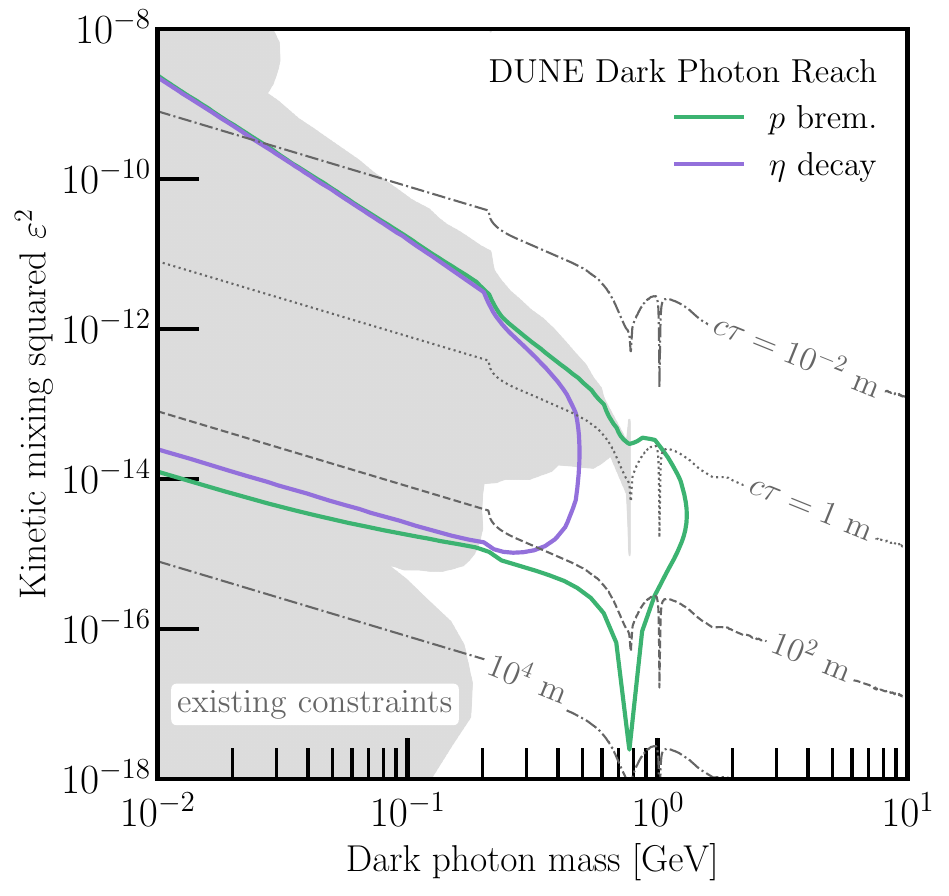}
\caption{Approximate reach of DUNE to discover model-specific long-lived particles, dark Higgs (left) and dark photons (right), assuming ten years of data collection. Sensitivity from different production mechanisms corresponds to different color contours -- the left panel shows charged kaon decay (blue) and proton bremsstrahlung of a scalar boson (red), where the right panel shows neutral $\eta$ meson decay (purple) and proton bremsstrahlung of a vector boson (green). Contours of lifetimes of the particles are shown by various gray lines, and gray shaded regions indicate parameter space excluded by existing searches~\cite{Batell:2022dpx,Gori:2022vri}.\label{fig:ModelSpecificSensitivity}}    
\end{center}
\end{figure}

The left panel of~\cref{fig:ModelSpecificSensitivity} demonstrates the parameter space of interest here; for Dark Higgs masses between $\SI{10}{MeV}$ and $\SI{10}{GeV}$, the particle is long-lived by neutrino-experiment standards and has sizeable decay branching ratios into final states of interest. Here, the gray, shaded regions indicate parameter space already ruled out by experiments~\cite{BNL-E949:2009dza,NA62:2020pwi, NA62:2021zjw, Foroughi-Abari:2020gju, ICARUS:2024oqb}.
We show a rough estimate of DUNE model-specific reach in this parameter space, using two dominant production modes that we will discuss in detail in the following: charged-kaon decay (blue) and proton bremsstrahlung (red).
Finally, as a basis for comparison, we show isocontours of dark Higgs lifetimes ranging from $\SI{1}{cm}$ to $\SI{e4}{m}$ -- the parameter space that DUNE will probe with these different production mechanisms spans a wide range of expected lifetimes for the dark Higgs.

%%%%%%%%%%%%%%
\textbf{Dark photon}
As a second benchmark scenario, we consider the dark photon model which introduces a new $U(1)_D$ gauge symmetry and associated gauge boson $A'$ that kinetically mixes with the SM photon, thereby acquiring couplings to charged SM particles \cite{Okun:1982xi, Holdom:1985ag, deNiverville:2011it, Berlin:2018pwi}.
The couplings strength is $\varepsilon q$, where $\varepsilon$ is the kinetic mixing parameter and $q$ is the electric charge of the SM particle. 
If $\varepsilon \ll 1$, as required by experimental constraints, these interactions are feeble, and in the absence of any other interactions $A'$ tends to be long-lived.
Its lifetime depends on both its mass and the kinetic mixing parameter, similar to the dark-Higgs scenario.

In a proton beam-dump environment, a number of production mechanisms are possible; which one dominates depends on the dark photon mass, $m_{A'}$.
For $m_{A'}$ below a few-hundred~MeV, production predominantly occurs via two-body neutral meson decays, e.g., $\pi^0 \to \gamma A^\prime$.
In contrast, above the $\eta$ meson mass, production via proton-proton bremsstrahlung, $p p \to p p A^\prime$, typically dominates. 
Nevertheless, even in the mass range where $\eta$ decays dominate, proton bremsstrahlung provides a comparable sensitivity, since the difference in production rates between the two mechanisms is less than an order of magnitude, and the more forward-boosted dark photons from bremsstrahlung experience reduced geometric suppression in the detector \cite{Berryman:2019dme}. 

The right panel of~\cref{fig:ModelSpecificSensitivity} demonstrates similar information for the dark photon as the left panel does for the dark Higgs, emphasizing sensitivity stemming from production via $\eta$ decay (purple) and proton bremsstrahlung (green) of the dark photon.
DUNE sensitivity extends beyond the reach of existing searches~\cite{NA482:2015wmo, LHCb:2019vmc, BaBar:2014zli, BaBar:2016sci, Chang:2016ntp, Bauer:2018onh,Tsai:2019buq}, largely due to the intensity of the proton beam and the relative detector position.
Again, we find that for the range of masses between approximately $\SI{10}{MeV}$ and $\SI{1}{GeV}$, these production mechanisms are relevant, and that a wide range of proper lifetimes of the long-lived particle can be tested in the DUNE facility.

\textbf{Leptophilic Gauge Boson}
Similar to the dark photon scenario, we will also consider models with gauge symmetries of the form $U(1)_{L_\alpha - L_\beta}$, where $\alpha,\ \beta \in \{e, \mu, \tau\}$ are lepton flavor indices \cite{He:1990pn, Foot:1994vd, Pospelov:2008zw}.
Models of this type are anomaly-free SM extensions whose phenomenology in beam-dump environments and elsewhere is qualitatively different from the dark Higgs and dark photon models~\cite{Bauer:2018onh}.
One phenomenological feature of gauged $L_\alpha - L_\beta$ models is the possibility of producing the new gauge boson, $V$, in three-body charged-meson decays such as $K^+ \to \mu^+ \nu_\mu V$, where $V$ is radiated off the charged-muon or muon-neutrino line in the final state.\!\footnote{Such three-body decays are possible in dark photon models but very often lead to smaller LLP yields than neutral-meson ($\pi^0$ and $\eta$) decays. In leptophilic gauge boson models, those neutral-meson decays depend on loop-induced kinetic mixing, where direct couplings to charged leptons/neutrinos can allow for relatively larger production rates.}
We do not display leptophilic gauge boson sensitivity for simplicity; we refer the interested reader to, for example, Refs.~\cite{Berryman:2019dme,Capozzi:2021nmp} for further discussion of this model in the context of DUNE.

\vspace{1ex}
In all three models (as well as others not considered here), a number of common features are present.
First, given the high energy of the protons striking the target, the $\sim $\,MeV--GeV LLPs produced will be highly boosted, focusing the LLPs in the forward direction.
This focusing is largely independent of the details of the production mechanism.
Moreover, in all three models the LLPs can decay into pairs of charged leptons, $\ell^+ \ell^-$, where $\ell = e, \mu$. 
A neutrino detector such as the DUNE ND is well-suited for studying such final states; calorimetric energy measurements along with shower/track reconstruction allow for the determination of the decaying particle's invariant mass.
But discriminating between $\ell^+ \ell^-$ from $\phi$, $A'$, and $V$ decay is disproportionately more difficult.

Note that other LLP scenarios are easier to tell apart. An example is a heavy neutral lepton (HNL) mixing with the SM neutrinos.
HNLs cannot decay into just an electron/positron or muon/antimuon pairs; only decays involving an extra final-state neutrino, such as ($e^+ e^- \nu$ and $\mu^+ \mu^- \nu$) are possible.
The lack of an peak in the $\ell^+ \ell^-$ invariant-mass distribution distinguishes HNL models from the scenarios considered here, as has been explored~\cite{Batell:2023mdn}.

The absolute signal rates expected in our three benchmark models depend strongly on the underlying parameters, in particular the LLP mass and mixing parameter.
However, given the unknown value of the mixing parameter, the total signal rate does not offer any model discrimination power, therefore we will focus mostly on the shape of kinematic distributions.
We will moreover use the number of signal events and the LLP lifetime as phenomenological variables.
In principle, using these two as well as the LLP mass introduces an extra degree of freedom relative to concrete models, where the only degrees of freedom are a mass and a coupling to the SM.
In these concrete scenarios, for a given mass, the coupling simultaneously determines the LLP lifetime and the expected event rate.
Allowing these two quantities to be separate allows for a more holistic, model-generic comparison (see, e.g., Refs.~\cite{Coloma:2019htx,Batell:2023mdn} for motivation of this approach).

%----------------------------------------------------------------------------------
\subsection{Simulation details}
%----------------------------------------------------------------------------------
We now describe how we simulate LLP production, propagation, and decay. In DUNE, mesons are produced in collisions of \SI{120}{GeV} protons with a fixed target. Our starting point are Monte Carlo event samples produced by the full DUNE beamline simulation~\cite{Fields,DUNE:2020ypp}, which contain in particular detailed kinematic information about charged mesons produced in the target.
The simulation accounts for the effect of the magnetic focusing horns, the production of secondary particles, and the propagation of mesons prior to their decay.
The DUNE simulation does not include neutral mesons (as they do not contribute to the neutrino flux); to understand LLP production from neutral-meson decay, we simulate their production using \texttt{PYTHIA-8}. Specifically, we generate sample meson fluxes by colliding \SI{120}{GeV} protons on at-rest target protons (using the flag \texttt{SoftQCD:all}), collecting the outgoing neutral mesons.
In doing so, we do not account for secondary processes, as this would require a more sophisticated simulated using (for instance) \texttt{GEANT-4}~\cite{GEANT4:2002zbu, Allison:2006ve, Allison:2016lfl}.
Including secondary production would lead to a slight increase in meson and LLP fluxes, so our approach is conservative.
We simulate the two- and three-body decays of mesons into LLPs directly\footnote{Two-body decays are simulated trivially in the meson's rest frame (isotropically) then boosted back to the laboratory frame. For three-body decays, we calculate the allowed Dalitz phase space for the decay and sample it uniformly, effectively ignoring the matrix element of these decays. Because the mesons are highly boosted, the impact of the matrix element should be negligible on observables related to the LLP flux downstream.}; proton-proton bremsstrahlung production of the scalar and vector LLPs is simulated using the formalism presented in Ref.~\cite{Boiarska:2019jym, Foroughi-Abari:2021zbm, Gorbunov:2023jnx}.
Once the LLP is produced, we track its propagation in the laboratory frame.

\begin{figure}
    \centering
    \includegraphics[width=0.95\linewidth]{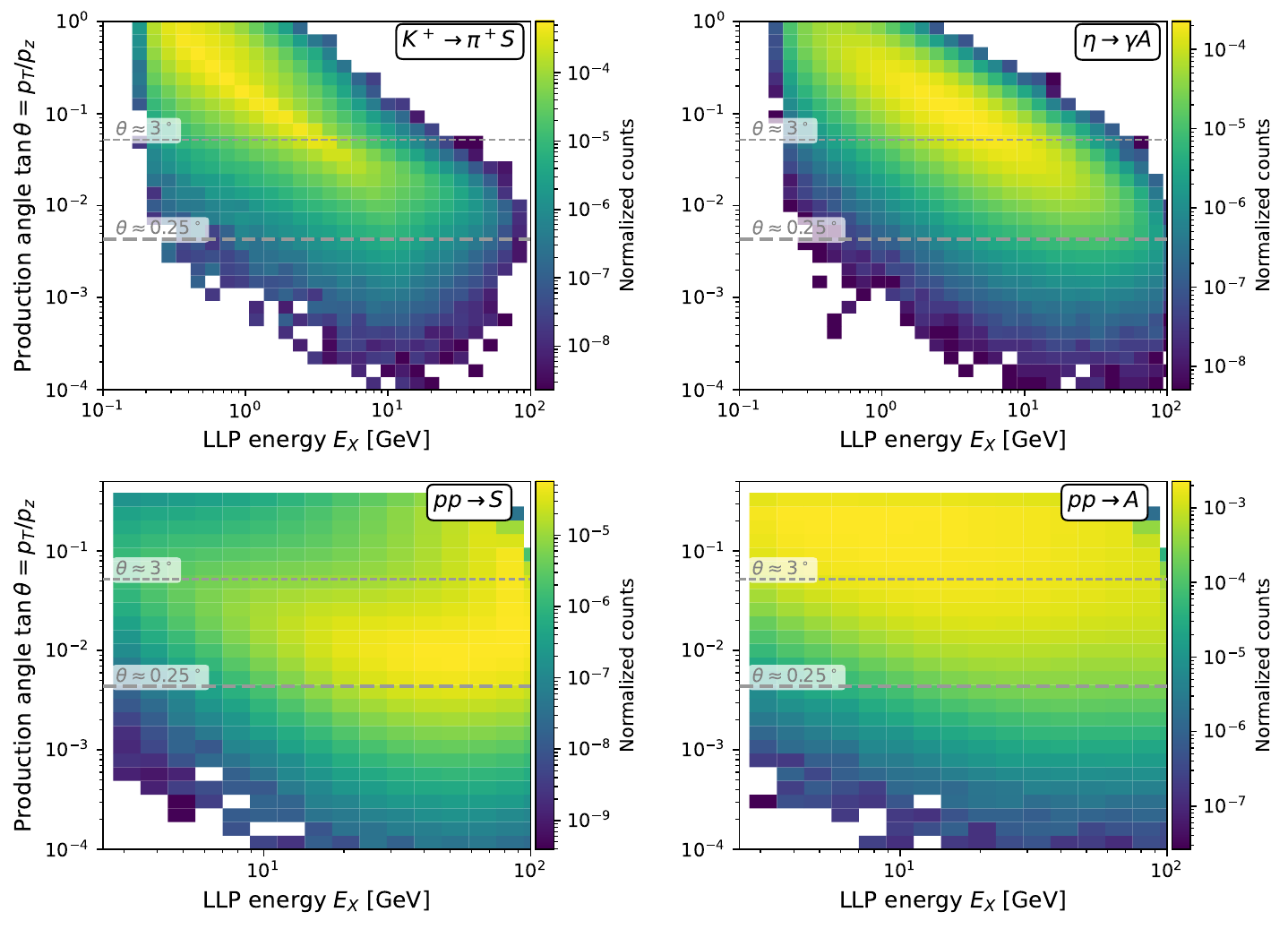}
    \caption{Kinematics of LLPs produced in the DUNE/LBNF beamline in the energy-vs-transverse momentum fraction plane).
    \label{fig:XProduction_Kaon_Eta}}
\end{figure}

Given that the main goal of this paper is to explore differences between LLP models, we begin in~\cref{fig:XProduction_Kaon_Eta} by showing the distribution of LLP energies, $E_X$, and production angles, $p_T/p_z$. Here, $X$ stands for $\phi$, $A'$, or $V$; $p_T$ and $p_z$ are the transverse and longitudinal components of the LLP momentum, respectively. In all cases, we have assumed an LLP mass of $m_X = \SI{200}{MeV}$. The DUNE ND covers approximately the range $p_T/p_z \lesssim 4 \times 10^{-3}$ when positioned on-axis.
The different panels in \cref{fig:XProduction_Kaon_Eta} correspond to different production mechanisms; the top panel shows two-body decays of kaons (left) and $\eta$ mesons (right), whereas the bottom panels show production via proton-proton bremsstrahlung of a scalar (left) or vector (right) LLP.

\begin{figure}
    \centering
    \includegraphics[width=\linewidth]{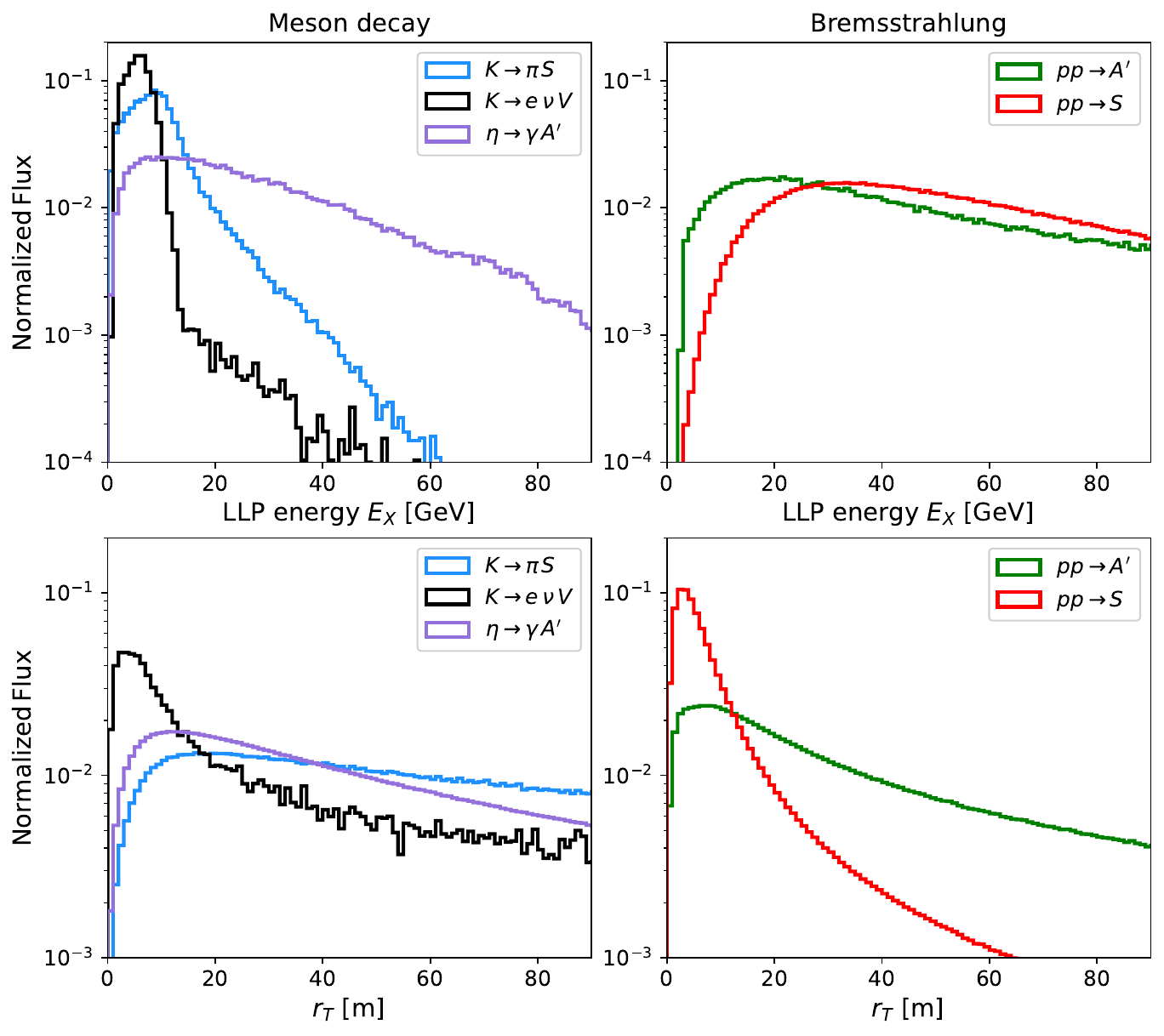}
    \caption{Top Panel : Area-normalized LLP fluxes at the DUNE near detector as a function of LLP energy. We show results for the dark Higgs ($S$), dark photon ($A'$), and leptophilic gauge boson ($V$), and for several different production channels (left panel: kaon decays, right panel: bremsstrahlung). In all cases, we have assumed an LLP mass of $m_X = \SI{200}{MeV}$. Bottom panel : Area-normalized fluxes with respect to transverse radius for five long-lived particle production methods, all of which having $m_X = 200$~MeV.
    \label{fig:Flux_2_body}}
\end{figure}

After accepting only those LLPs with small enough $p_T/p_z$ to intersect the DUNE ND in the on-axis position, we may inspect the LLP energy distribution. This is presented in~\cref{fig:Flux_2_body}(top), again for four production mechanisms all with $m_X = \SI{200}{MeV}$. 
We note here that the fluxes shown in \cref{fig:Flux_2_body} are all area-normalized.
We see that proton--proton bremsstrahlung yields higher-energy LLPs than meson decay due to the collinear divergence of bremsstrahlung production.
Comparing different meson decays, we note that LLPs from $\eta$ decay are higher in energy than those from kaon decay.
This is because electrically neutral $\eta$s are not focused, so only LLPs from the decays of the highest-energy $\eta$s are sufficiently boosted in the forward direction to intersect the detector.
LLPs from $K^\pm$, on the other hand, are forward-focused also at lower energies thanks to the magnetic horns.

\begin{figure}
    \centering
    \includegraphics[width=0.8\linewidth]{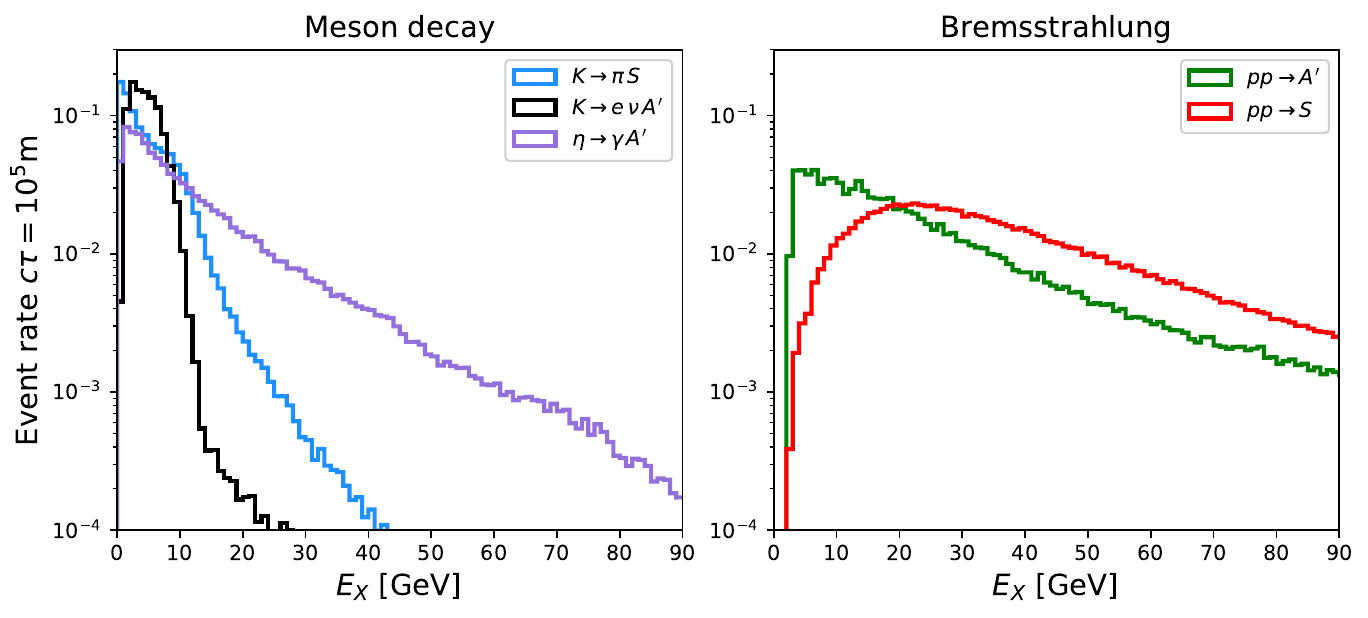}
    \caption{Normalized event rates with respect to particle energy for five long-lived particle production methods, all of which having $m_X = 200$~MeV, where we fix the lifetime to be $c\tau = 10^5$ m. The left panel shows meson-decay production: two-body decay of a charged kaon (blue), two-body decay of an $\eta$ meson (purple), and three-body radiative decay of a charged kaon (black). The right panel contrasts two proton-bremsstrahlung production modes: that of a vector (green) vs.~that of a scalar (red). For comparison, a smaller-lifetime ($\SI{10}{m}$) version of the expected event-rate distrubtion is shown in~\cref{fig:Event_rate_2_body_shortlifetime}.}
    \label{fig:Event_rate_2_body}
\end{figure}

While the fluxes shown in \cref{fig:XProduction_Kaon_Eta,fig:Flux_2_body} are instructive, they cannot be directly translated into event rates in the detector.
This is because the probability of the LLP decaying inside the detector is a nontrivial function of the LLP energy.
In the limit that the LLP lifetime is much longer than the distance to the detector, this probability becomes proportional to $m_X/E_X$, favoring decays of less boosted particles.
For shorter rest-frame lifetimes, more boosted particles will instead have a higher likelihood of decaying inside the instrumented detector volume.
In~\cref{fig:Event_rate_2_body}, we depict how this effect is incorporated for the fluxes shown in~\cref{fig:Flux_2_body}, now including the probability of the decay happening in the detector when the LLP lifetime is $c\tau = \SI{e5}{m}$.
An alternative, shorter-lifetime ($\SI{10}{m}$) event-rate distribution is presented in the appendix in~\cref{fig:Event_rate_2_body_shortlifetime}.

We are interested in going beyond purely energy-based observables to motivate better discrimination between these different LLP production modes.
To that end, we also incorporate spatial information, specifically the (transverse) distance between the beam axis and the decay vertex inside the detector -- $r_T = \sqrt{x^2 + y^2}$ (assuming the beam axis is in the $+z$-direction).
Considering this radial information, we allow our analysis to extend to a larger radial geometry, allowing for a combined analysis of event distributions with respect to \textit{simultaneously} the deposited energy and the vertex of the event.
This approach allows us to emulate the scenario where the detector is positioned off-axis, akin to the proposed DUNE-PRISM setup.
We see the radial distribution of events (in a hypothetically large detector volume) up to tens-of-meters of transverse radius in~\cref{fig:Flux_2_body}(bottom) -- many of the qualitative takeaways we understood from flux distributions are present here for the different production modes.
We will demonstrate how this additional information adds to our statistical separation between production mechanisms in the following -- the differentiation arises strongly from the differences between the parent production distributions, e.g., the difference in focusing between charged kaons and neutral etas.

\textbf{Statistical approach -- }
Thus far, we have relied on visual inspection of one-dimensional histograms to distinguish among production modes.
To provide a more statistically rigorous comparison, we perform a $\chi^2$-based analysis. 
We choose a particular `truth' hypothesis (e.g., a LLP with mass $200$~MeV, produced via charged kaon decay, with a lifetime of $100$~m) and simulate an expected event distribution according to this scenario.
In general, we will fix the total expected event rates to be $10$~or $100$~signal events in a given hypothesis -- in typical scenarios, this will be suitable to establish discovery in DUNE without having already been detected by other facilities~\cite{Berryman:2019dme,Coloma:2020lgy,Kelly:2020dda,Brdar:2020dpr,Dev:2021qjj,Coloma:2023oxx,Brdar:2025hqi}.
We then compare this hypothesis against a range of test hypotheses, allowing the test lifetime to vary between $10^{-2} - 10^{7}$~m.
In general, we will study discrimination between two concrete models, e.g.~the truth hypothesis $X$ and a test hypothesis $Y$.
We construct test-hypothesis event-rate distributions as a function of model-fractions $f_X$ and $f_Y$, fixing $0 \leq f_{X,Y} \leq 1$ and $f_Y = 1 - f_X$.
The true model discrimination then, is presented as a $\Delta \chi^2$ between the $f_X = 0$ and $f_X = 1$ points, marginalizing over the test lifetime.

Concretely, we will compare this type of model discrimination in two situations -- one in which the $\chi^2$ is calculated binned with respect to the LLP energy distribution, and one where (additionally) the transverse-displacement observable is included (in the range of the on-axis DUNE ND facility), yielding a two-dimensional $\chi^2$ comparison.
In both cases, we will further assume that the signal-to-background ratio is large and that the invariant mass of the LLP is measured precisely enough (by reconstructing the fully visible final state) so that in comparing production hypotheses, the mass does not need to vary.

The event expectation is calculated in terms of the expected energy spectrum, as well as the off-axis position (measured by $r_T$) in cases where we consider the spatial informaiton simultaneously in terms of $2$D distributions. The distributions are binned with respect to energy between $0$ and $100$~GeV in $5$~GeV bins, and with respect to off-axis position between $0$~and $2.5$~m in $25$~cm bins. These bin sizes are relatively conservative compared with what the DUNE liquid argon near detector should be able to accomplish, however we do not expect that finer binning will improve the model-by-model discrimination substantially.

Finally, we will analyze the simulated data assuming that background contributions can be ignored. A great deal of phenomenological effort has explored the possibility of reducing SM backgrounds (e.g., from neutrino scattering) to various LLP signatures, especially final states with electrons and positrons in Refs.~\cite{Coloma:2023oxx,Brdar:2025hqi}.
These strategies often take advantage of differences in the kinematic properties of LLP-decay signatures vs. neutrino-scattering backgrounds, e.g., targeting specific invariant masses of final-states or the delay of a (relatively massive) LLP arriving in the detector after the (nearly massless) neutrinos, e.g., Refs.~\cite{Shrock:1978ft, FMMF:1994yvb, T2K:2019jwa, MicroBooNE:2019izn}.
We refer the interested reader to these works for more detail.

%==================================================================================
\section{Analysis and results}
\label{sec:Results}
%==================================================================================
In this section, we present a number of pairwise comparisons between different truth- and test-scenario hypotheses, determining the ability to utilize energy and position information to differentiate between the two. First (in~\cref{subsec:ResultsTwoBody}), we compare among `two-body' production mechanisms, specifically two-body meson decays as well as proton/proton bremsstrahlung of LLPs. Then, motivated by leptophilic gauge bosons, we additionally compare against three-body meson decays as a production mode in~\cref{subsec:ResultsThreeBody}.

%----------------------------------------------------------------------------------
\subsection{Comparisons between Two Production Mechanisms}
\label{subsec:ResultsTwoBody}
%----------------------------------------------------------------------------------
In these analyses, we will compare two different hypothetical production mechanisms, taking one as the assumed-true scenario and the other as a test hypothesis.
For these comparisons, we will assume the total number of expected events is $10$, and expect that the sensitivity will scale roughly linearly with this (so a factor of ten larger $\chi^2$ discrimination if the number of signal events is $100$, for instance).
We will allow the fraction attributed to each of the two production mechanisms to vary (between $0$ and $1$), fixing the total expected events summed over the two mechanisms to $10$.

\textbf{Two-Body Decay -- Kaon vs.~Eta Production:} As a first comparison, we choose the two, two-body meson decay channels discussed above: decays of $K^\pm \to \pi^\pm X$ and $\eta \to \gamma X$.
%%%%%%%
\begin{figure}[!htbp]
\centering
\includegraphics[width=\linewidth]{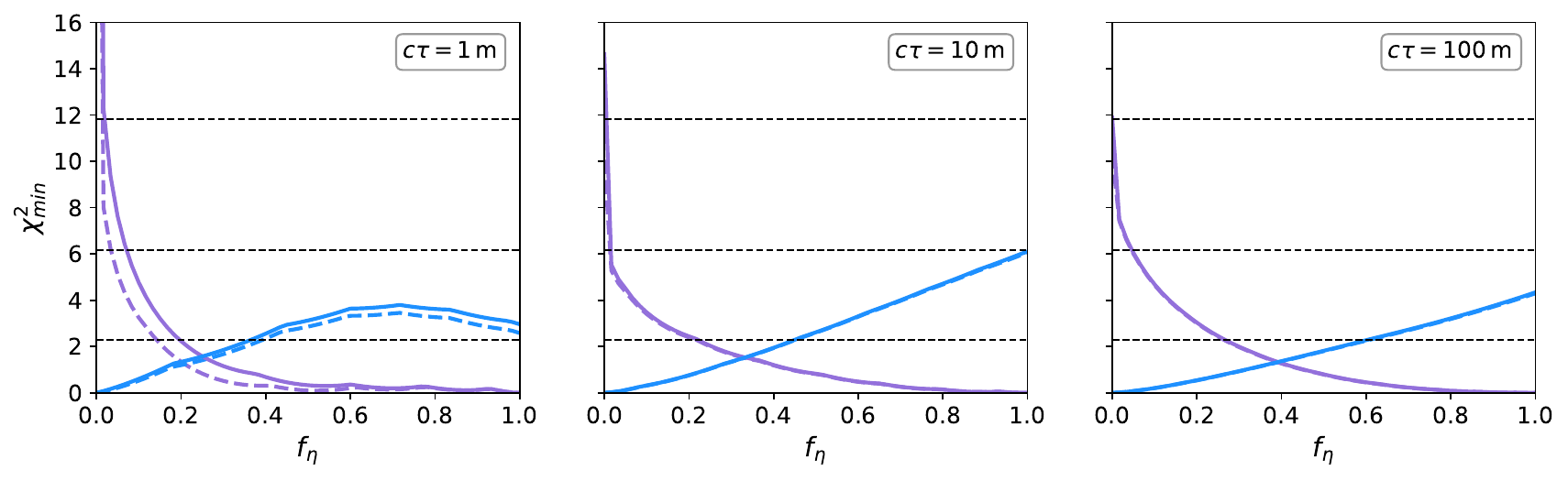}
\caption{Minimum likelihood comparison between the channels $K^\pm \to \pi^\pm S$ and $\eta \to \gamma A'$. The horizontal axis shows the assumed production fraction from $\eta$ decays, while the vertical axis gives the minimum $\chi^2$ value obtained when testing a hypothesis against the true one. Curves in purple correspond to the case where the true hypothesis is $\eta$ decay, and curves in blue correspond to kaon decay. Solid lines indicate a two-dimensional comparison that includes both energy and radial distributions, whereas dashed lines represent a one-dimensional comparison based only on the energy spectrum. The three dashed horizontal lines corresponds to values of $1\sigma, \,2\sigma$ and $3\sigma$ for reference. The 3 panels corresponds to different choices of LLP lifetimes, viz., 1 m, 10 m and 100 m. }
\label{fig:ChiSq_Kaon_Eta}
\end{figure}
%%%%%%%%map
We choose three representative `assumed-true' LLP lifetimes, $1$~m, $10$~m, and $100$~m\footnote{Since the distance between production and the detector is approximately $500$~m and accounting for the effect of the particles' boosts, lifetimes larger than ${\sim}100$~m look roughly identical to the $100$~m case.}, and simulate event-rate distributions assuming production via either the $K^\pm$ or $\eta$ decay process.
We then fit the data, assuming a test hypothesis of an admixture of the two production modes and a test lifetime, arriving at a chi-squared comparison. 
For a first analysis, we follow this process using the energy distributions of the LLPs only.
These $\chi^2$ values, as a function of the fitted $\eta$ fraction $f_\eta = 1-f_K$, are shown in~\cref{fig:ChiSq_Kaon_Eta} for the three assumed-true lifetimes.
In each panel, the blue (purple) lines correspond to the case where the true production mechanism comes from $\eta$ ($K^+$) decay.
If we additionally include transverse-radius information of the event rates (within the on-axis detector), we arrive at the solid lines in each panel -- the improvement attained by simultaneously including energy- and position-information is represented by the gap between the dashed and solid lines.
As is evident, whether this improvement is significant depends on the truth hypothesis (including the assumed-true lifetime) at play.

In each case, a relevant figure of merit is the potential of ruling out the `wrong' production mode, i.e.~ruling out $f_\eta = 1$ when the true production is from kaon decay and vice versa.
Generically, we see that it is easier to rule out kaon production when the truth is eta production than the opposite case -- this is related to the fact (see~\cref{fig:Flux_2_body}) that the LLPs from $\eta$ decay, passing through DUNE ND, tend to be more energetic than those produced from $K^+$ decay.
As a result, it is more challenging for a distribution of $K$-produced LLPs to mimic that of $\eta$-produced ones than vice versa, even when allowing the lifetime of the test hypothesis to vary.
\textit{Model-specific interpretation:} if a dark photon exists and DUNE discovers it through $\eta$-decay production, we expect that we can confidently identify the new-physics particle as a dark photon instead of being a dark Higgs boson. The converse is not true: if a dark Higgs is discovered through $K^\pm$ decay, interpreting its nature will prove challenging.

\begin{figure}[!htbp]
\centering
\includegraphics[width=\linewidth]{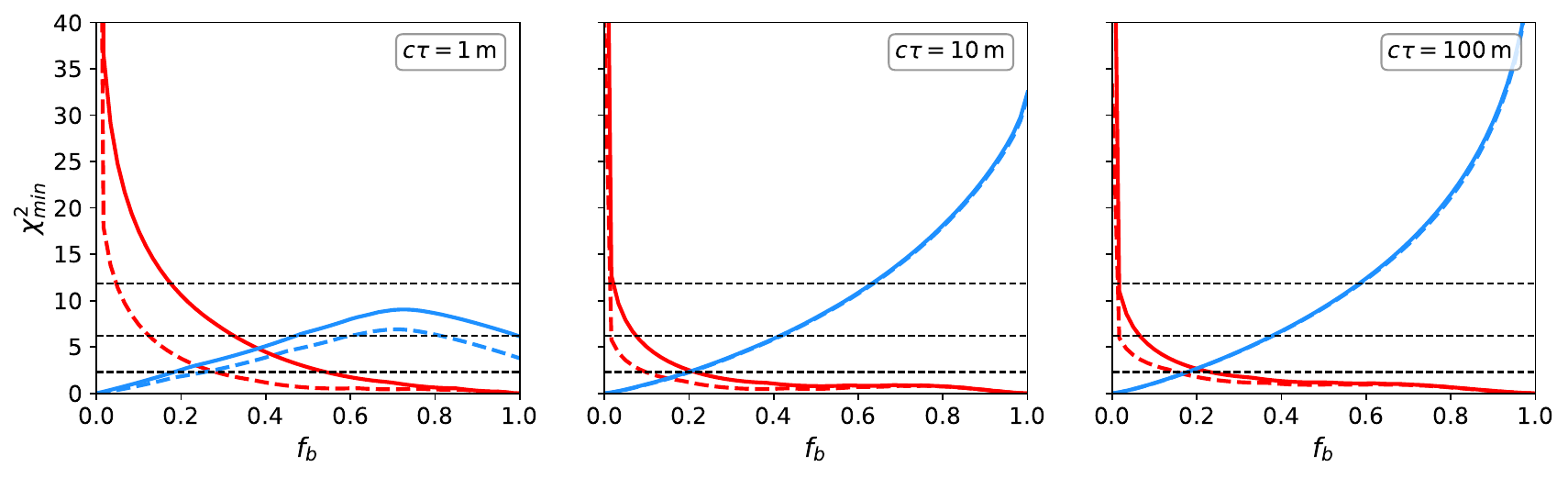}
\caption{Minimum likelihood comparison between the channels $pp \rightarrow S$ vs $K^\pm \rightarrow \pi^\pm S$. The horizontal axis shows the assumed production fraction from proton Bremsstrahlung, while the vertical axis gives the minimum $\chi^2$ value obtained when testing a hypothesis against the true one. Curves in red correspond to the case where the true hypothesis is proton Bremsstrahlung, and curves in blue correspond to kaon decay.\label{fig:chisq_brem_k}}
\end{figure}

%%%%%
\textbf{Scalar Production -- Kaon-Decay vs.~Bremsstrahlung:} For a second comparison, we consider two disparate production mechanisms of new scalar bosons: that from charged-kaon decay (blue) vs. from proton bremsstrahlung (red).
We use $f_b$ to represent the production fraction according to bremsstrahlung and take $f_K = 1-f_b$.
The results of this $\chi^2$ analysis are presented in~\cref{fig:chisq_brem_k}. Especially for the small-lifetime case ($1$~m, left panel), we see that the inclusion of radial information has a significant impact.
This is especially true when the truth hypothesis is bremsstrahlung production.

\textit{Model-specific interpretation:} if a dark Higgs boson is discovered, it will be relatively straightforward (except in specific cases) to differentiate between whether it is dominantly produced via $K^\pm$ decay or proton bremsstrahlung. If the LLP's lifetime is small, the inclusion of $r_T$ observations and the operation of DUNE-PRISM will aid substantially in this differentiation.

\begin{figure}[!htbp]
\centering
\includegraphics[width=\linewidth]{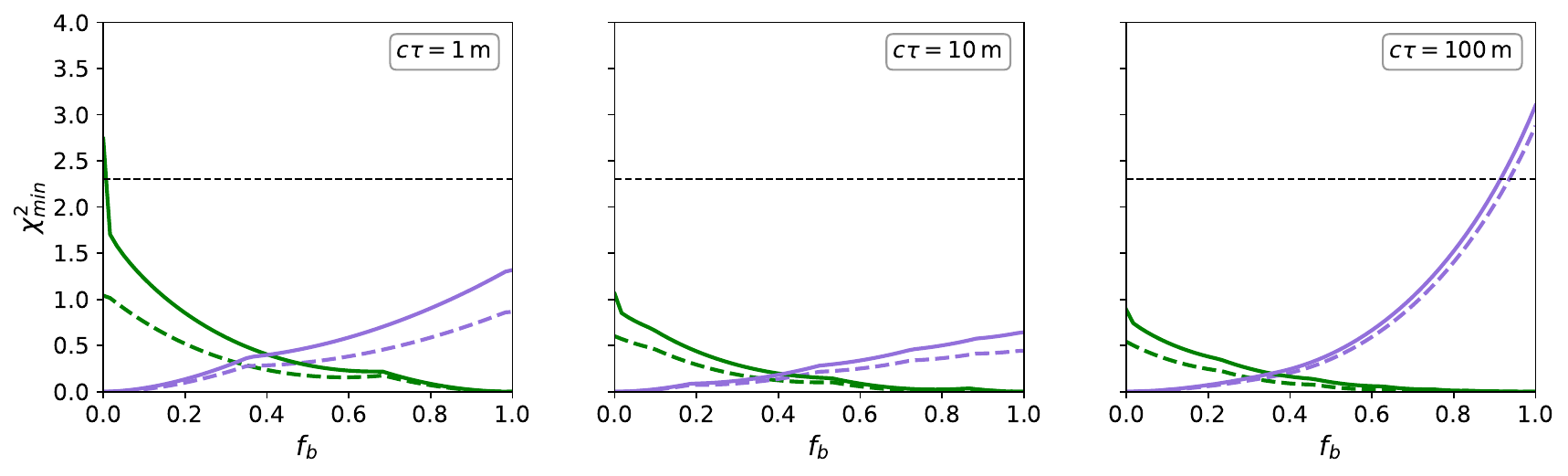}
\caption{ Minimum likelihood comparison between the channels $pp \rightarrow A'$ vs $\eta \rightarrow \gamma A'$. The horizontal axis shows the assumed production fraction from proton Bremsstrahlung, while the vertical axis gives the minimum $\chi^2$ value obtained when testing a hypothesis against the true one. Curves in green correspond to the case where the true hypothesis is proton Bremsstrahlung, and curves in purple correspond to eta decay.\label{fig:chisq_brem_eta}}
\end{figure}

%%%%%%%
\textbf{Vector Production -- Eta-Decay vs.~Bremsstrahlung:} For production of vector-type LLPs, two common modes include neutral $\eta \to \gamma A^\prime$ decays and proton-bremsstrahlung $p p \to A^\prime$. We show the results of a differentiation-analysis between these two in~\cref{fig:chisq_brem_eta}, where the purple (green) lines correspond to the truth hypothesis being $\eta$ decay (proton bremsstrahlung). The production fractions are defined as $f_b$ for the bremsstrahlung production fraction and $f_\eta = 1-f_b$ for the $\eta$ decay.

We note that the vertical scale in~\cref{fig:chisq_brem_eta} is suppressed relative to previous figures, demonstrating that differentiating between these two production modes with $\mathcal{O}(100)$ signal events (even with optimistic assumptions regarding background and reconstruction techniques) is extremely challenging. Even though adding in vertex position information (the increase from dashed to solid lines) modestly improves the $\Delta\chi^2$, these channels remain very challenging to disentangle. \textit{Model-specific interpretation:} If we know that a new-physics signal is associated with a dark photon decay, it will prove very difficult to verify how the dark photons are produced.

%%%%%%
%%%%%%%%%

\begin{figure}[!htbp]
\centering
\includegraphics[width=\linewidth]{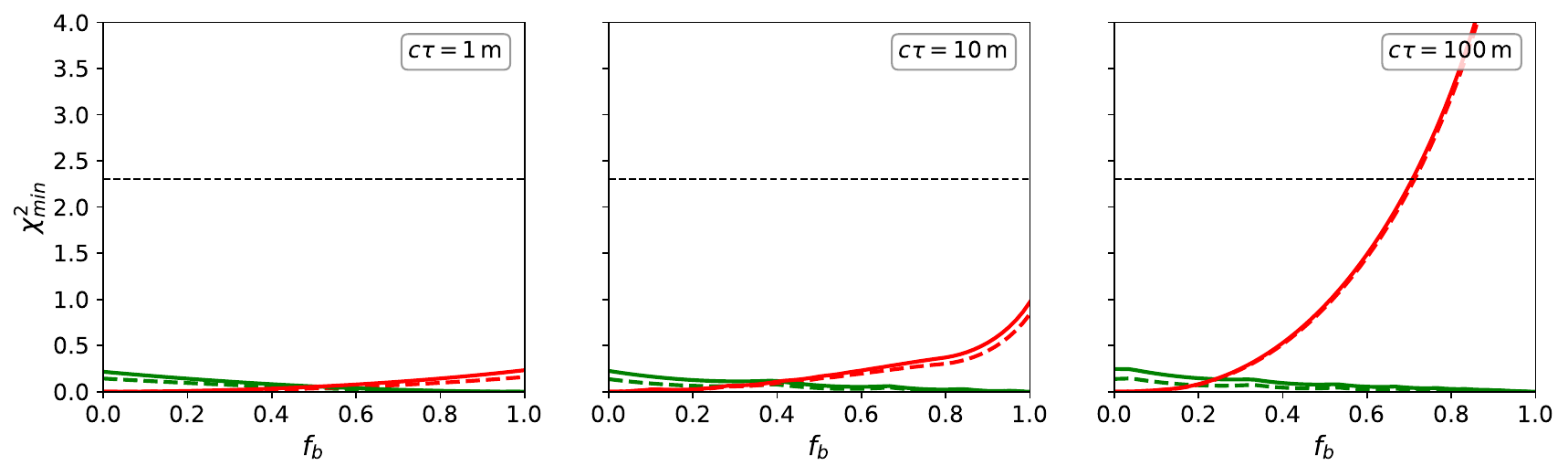}
\caption{Minimum likelihood comparison between the channels $pp \rightarrow S$ vs $pp \rightarrow A'$. The horizontal axis shows the assumed production fraction from proton Bremsstrahlung to dark photon, while the vertical axis gives the minimum $\chi^2$ value obtained when testing a hypothesis against the true one. Curves in green correspond to the case where the true hypothesis is proton Bremsstrahlung to dark photon, and curves in red correspond to proton Bremsstrahlung to scalar.\label{fig:chisq_brem_sc_g}}
\end{figure}

\textbf{Bremsstrahlung comparison -- Scalar vs.~Vector:} As a final direct comparison between production hypotheses, we consider production purely via bremsstrahlung in an attempt to differentiate between the LLP being a scalar (red) vs.~a vector (green) particle. This is motivated in many LLP situations for mass ranges where such bremsstrahlung production is the dominant expectation but there is no a priori reason to expect a vector- vs.~scalar-LLP. We show the results of the comparison in~\cref{fig:chisq_brem_sc_g}.
As with the $\eta$-decay vs.~bremsstrahlung case above, we see that the channels are extremely challenging to differentiate. The only likely hope in doing so would come from an analysis that can measure to significantly larger $r_T$, as exemplified in the difference between the red and green lines of~\cref{fig:Flux_2_body}(bottom). This should be possible with the DUNE-PRISM concept~\cite{DUNE:2025lvs}, however it may be challenging to collect sufficient statistics to differentiate between the two situations.

%----------------------------------------------------------------------------------
\subsection{Model-inspired Comparison of Three Production Mechanisms}
\label{subsec:ResultsThreeBody}
%----------------------------------------------------------------------------------
Here, in contrast to the above, we focus on a specific comparison between three production mechanisms, inspired by models of dark gauge bosons.
We simultaneously consider production via proton bremsstrahlung, decays of $\eta$ mesons, and the three-body decay from charged kaons, $K^+ \to \ell^+ \nu V$.
Such production mechanisms could all play a comparable role for detection in an experiment such as DUNE in, for instance, models where $L_\mu - L_\tau$ lepton number is promoted to a gauge symmetry.
For the following, we assume that the mass of the LLP is fixed to $200$~MeV and that we can detect $100$~signal events (in contrast to the 10-signal-event expectation in the previous subsection).
We consider potential truth lifetimes of $1$ and $100$ meters, but allow the test lifetime to vary independently in our analyses.

To allow for three production modes to vary simultaneously we define scaling parameters $f_b$, $f_\eta$, and $f_K$ (corresponding to the modes listed above), but fix $f_b + f_\eta + f_K = 1$. In our comparisons, we find that the inclusion of both energy- and spatial-information ($E_X$ and $r_T$ distributions, respectively) provides $\mathcal{O}(1)$ improvement over energy-based analyses alone, without significantly changing the main takeaways of such analyses. In~\cref{app:1D2DComparison} we compare the results of such one-dimensional and two-dimensional analyses; that aside, the remainder of this subsection focuses on two-dimensional analyses with both energy and spatial information included.

\begin{figure}[!htbp]
    \centering
    \includegraphics[width=\linewidth]{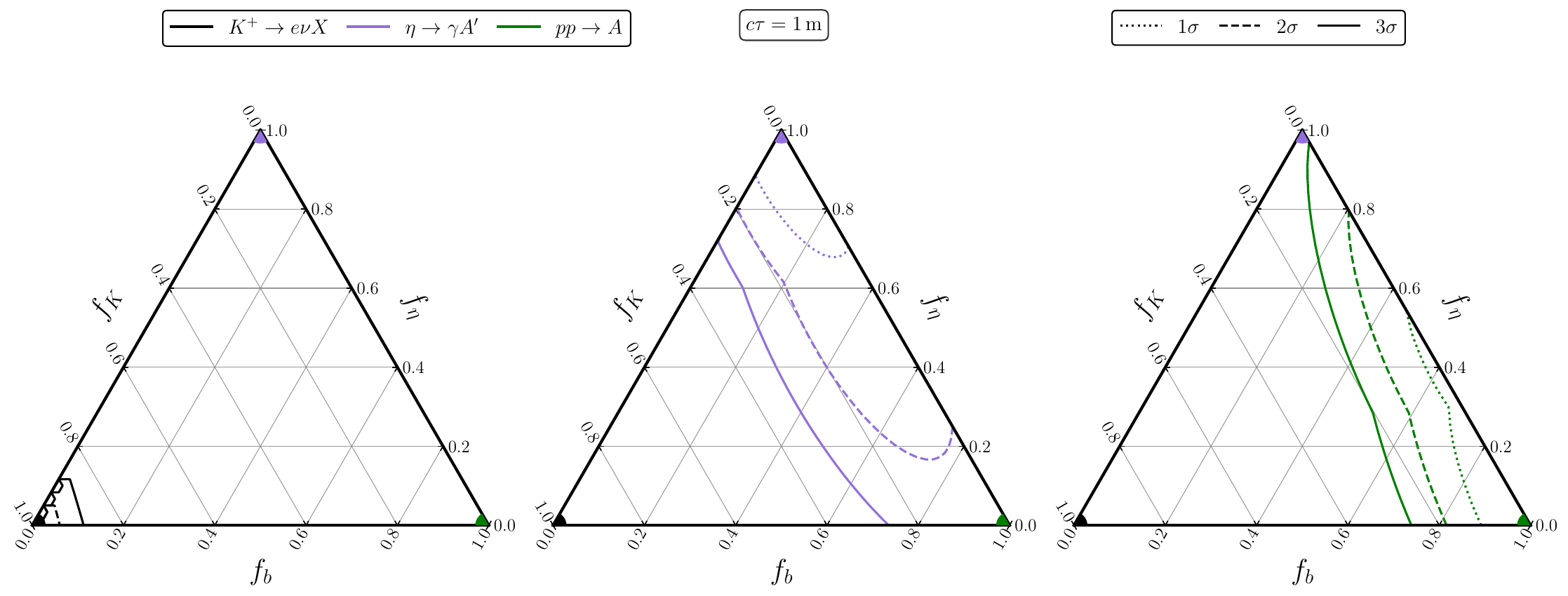}
    \vspace{0.5cm} % space between rows

    \includegraphics[width=\linewidth]{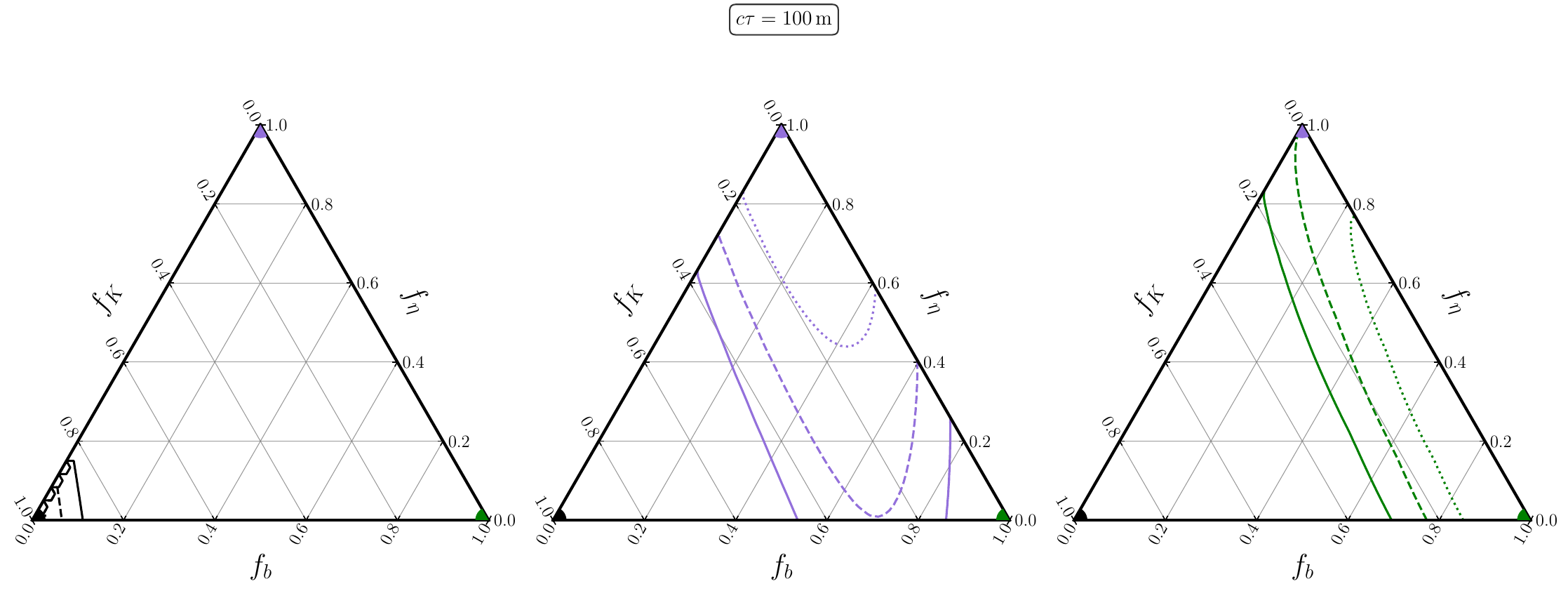}
    \caption{Minimum likelihood comparison among production hypotheses: $\eta \rightarrow A'$, $pp \rightarrow A'$, and $K \rightarrow \nu_e\, e\, Z'$ for different values of the proper decay length \( c\tau \).}
    \label{fig:chisq_3_comp_all}
\end{figure}

To visualize the interplay between the three signal hypotheses, we employ ternary plots, as shown in Fig.~\ref{fig:chisq_3_comp_all}, for benchmark decay lengths of $c\tau = \{1,\,100\}\,\mathrm{m}$ (top and bottom rows, respectively). Each corner of the triangle represents a pure production mode: the left vertex corresponds to $f_K=1$, the right vertex to $f_b=1$, and the top vertex to $f_\eta=1$, with the production fractions summing to unity at every point:
\[
f_\eta + f_K + f_b = 1.
\]  
Contours in each triangle indicate regions of constant $\chi^2$ assuming one of the three channels as the true production mode. In both rows, the left-most triangle corresponds to kaon three-body decay, the central triangle to $\eta$ decay, and the right-most triangle to proton bremsstrahlung. Dashed and solid lines indicate different confidence levels -- $1\sigma$ and $2\sigma$ respectively -- for each hypothesis being tested against mixtures of all three.

The sharp localization near the kaon vertex (left panels, in black) illustrates its strong discriminatory power: even small admixtures of other channels lead to a poor fit. In contrast, the broader and flatter contours for the bremsstrahlung hypothesis (right panels, in green) indicate that it can be more easily mimicked by other channels, consistent with our earlier observation that $\eta$ and bremsstrahlung contributions are partially degenerate.  
Overall, the ternary representation provides a compact and intuitive visualization of the distinguishability among the three production mechanisms across the full range of possible mixtures, complementing the 1D and 2D $\chi^2$ analyses discussed in~\cref{app:1D2DComparison}.

As the lifetime increases to $c\tau = 100~\mathrm{m}$ (bottom row), the corresponding boost factor becomes larger and making the discrimination for the kaon three-body decay hypothesis slightly less robust. On the other hand, when the true hypothesis is $\eta$ decay (center panel, purple), it can still be imitated by proton bremsstrahlung, and vice versa: the bremsstrahlung hypothesis can also be mimicked by $\eta$ decay, although the discriminatory power for $\eta$ decay worsens whereas the impact on proton bremsstrahlung remains unchanged by the increase in lifetime. 

%%%%%%%%%%%%%%%%%%%%%%%%%%%%%%%%%%%%%%%%
\section{Conclusions}\label{sec:Conclusions}
%%%%%%%%%%%%%%%%%%%%%%%%%%%%%%%%%%%%%%%%
In this work we have explored the prospects for distinguishing different production mechanisms of long-lived particles (LLPs) at the DUNE near detector. 
Whereas most existing studies emphasize sensitivity to specific LLP models, our focus is on the complementary question of how well distinct production modes can be disentangled, in a statistical sense, once a signal is observed. 
We consider representative benchmark models---a dark Higgs, a dark photon, and anomaly-free $L_\alpha-L_\beta$ scenarios---and study their dominant production channels: 
(i) two-body meson decays, 
(ii) bremsstrahlung off the primary proton beam, and 
(iii) three-body meson decays. 
For meaningful comparisons, we fix both the signal event yield and the LLP mass when contrasting different channels. 
Our analysis employs pairwise comparisons of production mechanisms, allowing for the possibility of mixed production fractions, and is further extended to simultaneous three-channel comparisons visualized using ternary plots. 
To quantify discrimination power, we construct truth and test hypotheses using binned $\chi^2$ statistics based on both the visible energy and the transverse displacement of LLP decays within the DUNE detector.

Our results demonstrate that kinematic information is crucial for disentangling different LLP production mechanisms at DUNE. 
While the visible energy spectrum alone often leaves substantial degeneracies, the inclusion of transverse displacement can break these degeneracies and enable robust model separation even with modest signal statistics ($\mathcal{O}(10$--$100)$ events). 
In particular, we find that separation between kaon and $\eta$ channels, as well as between charged-meson decays and proton bremsstrahlung, improves significantly once spatial information is taken into account, underscoring the importance of exploiting the full DUNE-PRISM capabilities and accounting for the effect of magnetic horns on charged-meson momentum distributions. 
Moreover, distinguishing three-body meson decays---which are especially relevant for $U(1)$ gauge-boson mediator models---from either two-body meson decays or bremsstrahlung proves relatively straightforward. 
By contrast, discriminating LLPs produced via meson decays from those produced via proton bremsstrahlung remains challenging; in such cases, additional observables, such as timing information, may be necessary to achieve concrete separation.

The approach developed here is broadly applicable to LLP searches at fixed-target experiment facilities. 
By moving beyond discovery sensitivity and quantifying model discrimination, 
our work illustrates the critical role that detector geometry and kinematic observables will play in the interpretation of any future LLP signal. 
In future studies it will be valuable to incorporate more realistic background estimates, detector response, 
and extended model spaces, including scenarios with two or more accessible production channels. 
Taken together, our results strengthen the physics case for DUNE as a leading facility in exploring the nature of long-lived particles beyond the Standard Model, and underscore the discovery potential of near-detector facilities.

\textbf{Note added:} While this work has been reviewed and approved as a DUNE theory paper by the DUNE Collaboration, the results presented herein represent the views of the authors and not of the DUNE Collaboration as a whole.

%%%%%%%%%%%%%%%%%%%%%%%%%%%%%%%%%%%%%%%%
\section*{Acknowledgments}
%%%%%%%%%%%%%%%%%%%%%%%%%%%%%%%%%%%%%%%%
We are very grateful to Joachim Kopp for thorough discussions and collaboration during the early stages of this work.
We thank Bhaskar Dutta, Pedro Machado, Robert Shrock, and Davi de Assis Camargos for valuable feedback on an early draft of this manuscript.
The work of KJK and MR is supported in part by US DOE Award \#DE-SC0010813.

%%%%
\appendix
\section{Additional Kinematic Distributions:\texorpdfstring{\\}{} Extended Radial Distributions, Varying LLP Mass, Alternative LLP Lifetimes}

\begin{figure}[!htbp]
\centering
\includegraphics[width=\linewidth]{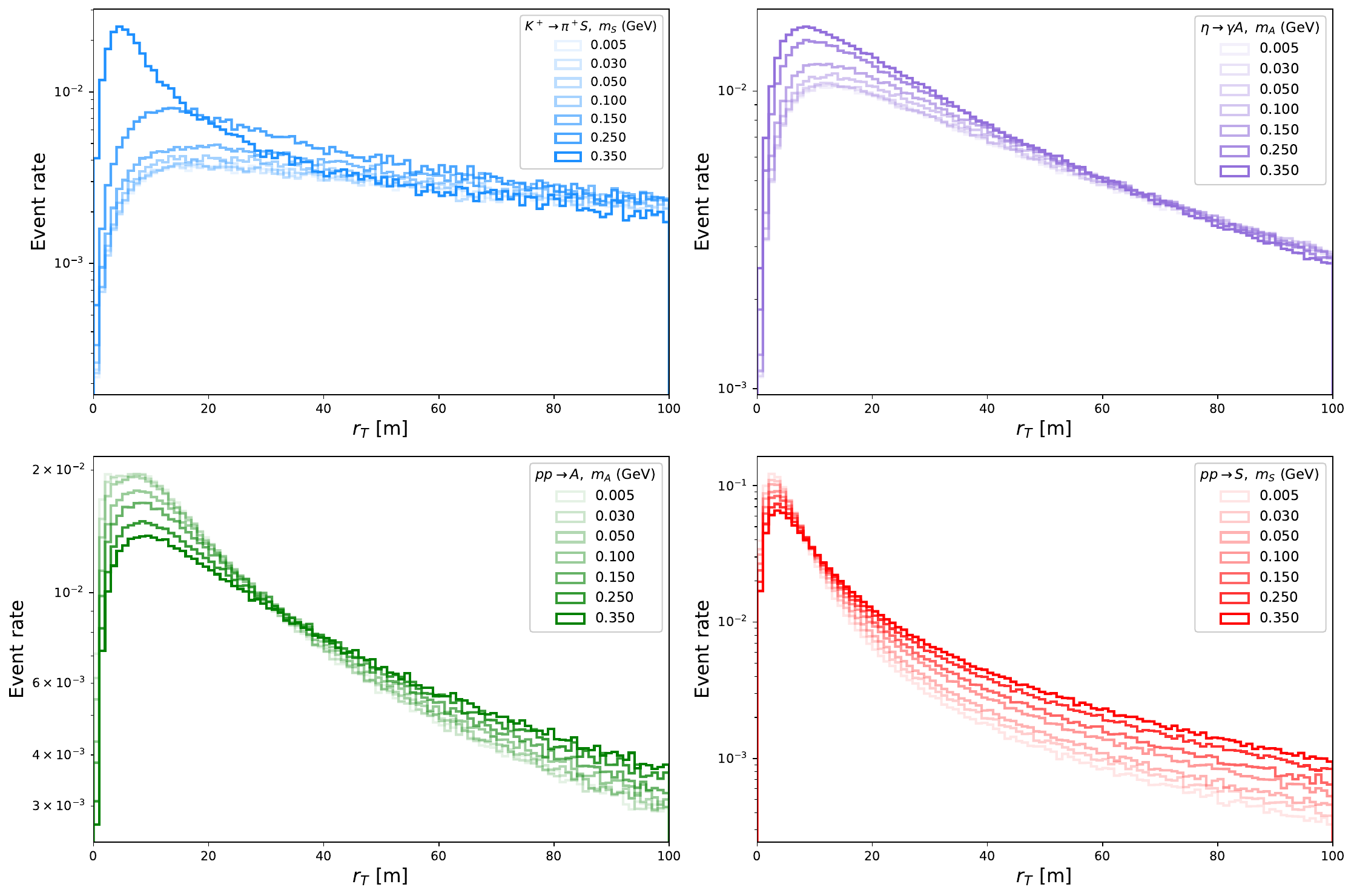}
\caption{Normalized event rates as a function of transverse radius with varying masses}
\label{fig:Transverse_mass_2_body}
\end{figure}

In analyzing expected LLP fluxes at the DUNE near detector, we constructed~\cref{fig:Flux_2_body}(bottom), which displays the transverse-radius distribution of various LLP production mechanisms. 
This was key in connecting the LLP detection to the DUNE-PRISM concept, where observations at small and large radii can assist in disentangling different production mechanisms and therefore different models of LLPs.
In doing so, we fixed our analysis to LLP masses of $\SI{200}{MeV}$ -- here, we repeat this procedure but allowing the LLP mass to vary.
The resulting fluxes, separated by production mechanism, are shown in~\cref{fig:Transverse_mass_2_body}: charged-kaon decay (blue, top-left panel), $\eta$ decay (purple, top-right), bremsstrahlung of a vector (green, bottom-left), and bremsstrahlung of a scalar (red, bottom-right).

We see that the LLP fluxes from charged-kaon decay have substantial mass-dependence, becoming more forward-peaked as the mass increases.
This is in contrast to the fluxes from $\eta$ decay, which remain more radially spread.
This difference arises due to the narrowing kinematical threshold as we approach the $m_K - m_\pi \sim \mathcal{O}(0.4\,\mathrm{GeV})$ window, in addition to the magnetic focusing of the charged kaons but not the neutral $\eta$ mesons.
In the bottom panels, we see that the mass-dependence of LLP fluxes from proton bremsstrahlung has similar characteristics when considering vector or scalar bosons, however the variance is not very large.

Additionally in~\cref{sec:Models}, we inspected expected event rates from different LLP production modes, assuming a long lifetime of $c\tau = \SI{e5}{m}$ (relative to the detector distance). We show a different, smaller choice of the lifetime for the same production mechanisms in~\cref{fig:Event_rate_2_body}
\begin{figure}[!htbp]
    \centering
    \includegraphics[width=0.8\linewidth]{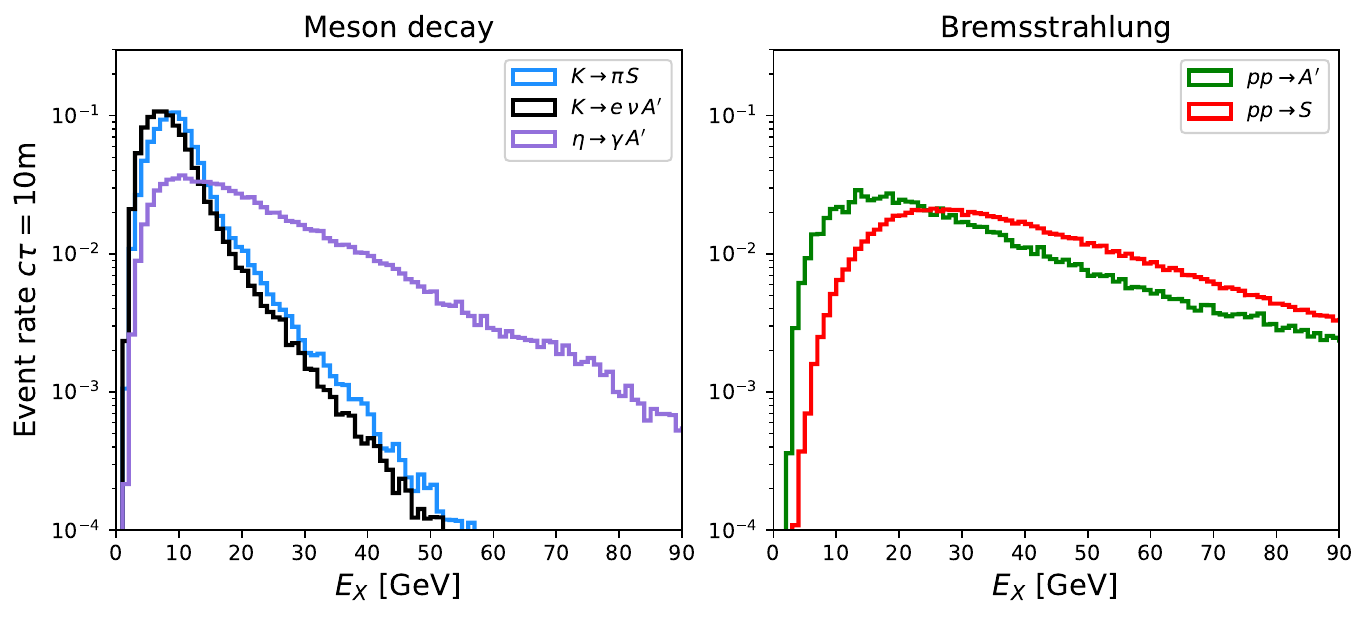}
    \caption{Expected event-rate distributions for different production mechanisms (see~\cref{fig:Event_rate_2_body} for a comparison), but with a shorter LLP lifetime of $c\tau = \SI{10}{m}$.\label{fig:Event_rate_2_body_shortlifetime}}
\end{figure}
Due to the shorter lifetime, LLPs with larger energies are favored to reach the detector and decay within -- we observe event-rate distributions shifted towards higher LLP energies as a result.

\section{Three-signal-hypothesis comparison with energy-/spatial-information}\label{app:1D2DComparison}
When considering three production mechanisms simultaneously in the analysis surrounding~\cref{fig:chisq_3_comp_all}, we focused on the situation in which we analyzed expected event rates in terms of LLP energy and transverse position simultaneously.
Here, we revisit these analyses, demonstrating the additional information gleaned by including transverse-radius $r_T$ information to the $\chi^2$ fit.

We will demonstrate this improvement by comparing side-by-side analysis results assuming either (a) in left panels, the measurement potential using \textit{only} LLP energy information, or (b) in the right panels, the measurement potential if $r_T$ information is \textit{additionally} incorporated.
We do so for different assumed-true production mechanisms, showing the three possibilities simultaneously in a given figure: three-body kaon decay (black), $\eta$ decay (purple), and proton bremsstrahlung (green), still restricting $f_K + f_\eta + f_b = 1$ in our analysis ($f_b$ is not shown for simplicity here; the upper-right triangular half of each panel is forbidden by this restriction).

\begin{figure}[!htbp]
\centering
\includegraphics[width=\linewidth]{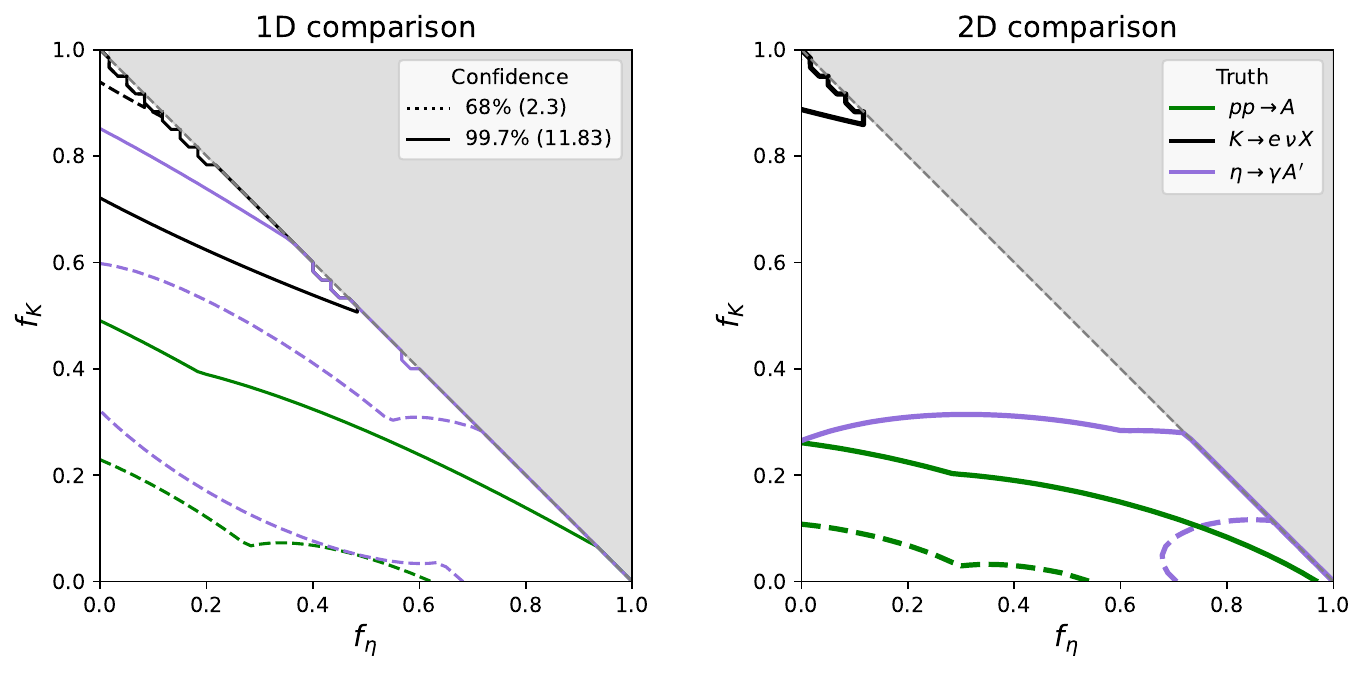}
\caption{Minimum likelihood comparison between the channels $\eta \to A'$, $pp \to A'$, and $K \to \nu_e, e, Z'$ for $c\tau = 100,$m. The horizontal axis indicates the production fraction from $\eta$ decays, while the vertical axis shows the production fraction from kaon decays. The left panel displays likelihood comparisons based solely on the energy distribution, whereas the right panel includes both energy and radial distributions. Solid lines denote the $3\sigma$ confidence contours, and dashed lines denote the $1\sigma$ contours. Black curves correspond to kaon three-body decays, purple to $\eta$ decays, and green to proton bremsstrahlung. The gray shaded region is excluded as the sum of production fractions exceed unity. Here the lifetime of LLPs is set to $c\tau =1$ m. \label{fig:chisq_3_comp_1m}}
\end{figure}
The first result is shown in~\cref{fig:chisq_3_comp_1m} for a true lifetime of $\SI{1}{m}$.
We see that the kaon hypothesis (black) produces the most distinct, small measurement contours: it is relatively easy to differentiate this production from the other mechanisms. Additionally, the contours do not change significantly going from the left to right panels: for this case, the radius-dependence does not add much more information.
In contrast, we see substantial improvement in the measurement contours for production via $\eta$ decay (purple) and proton bremsstrahlung (green).
For instance, this additional information provides additional means of separating the proton-bremsstrahlung and kaon-decay hypotheses, as evidenced by the green, solid line moving down from the left panel to the right one.

\begin{figure}[!htbp]
\centering
\includegraphics[width=\linewidth]{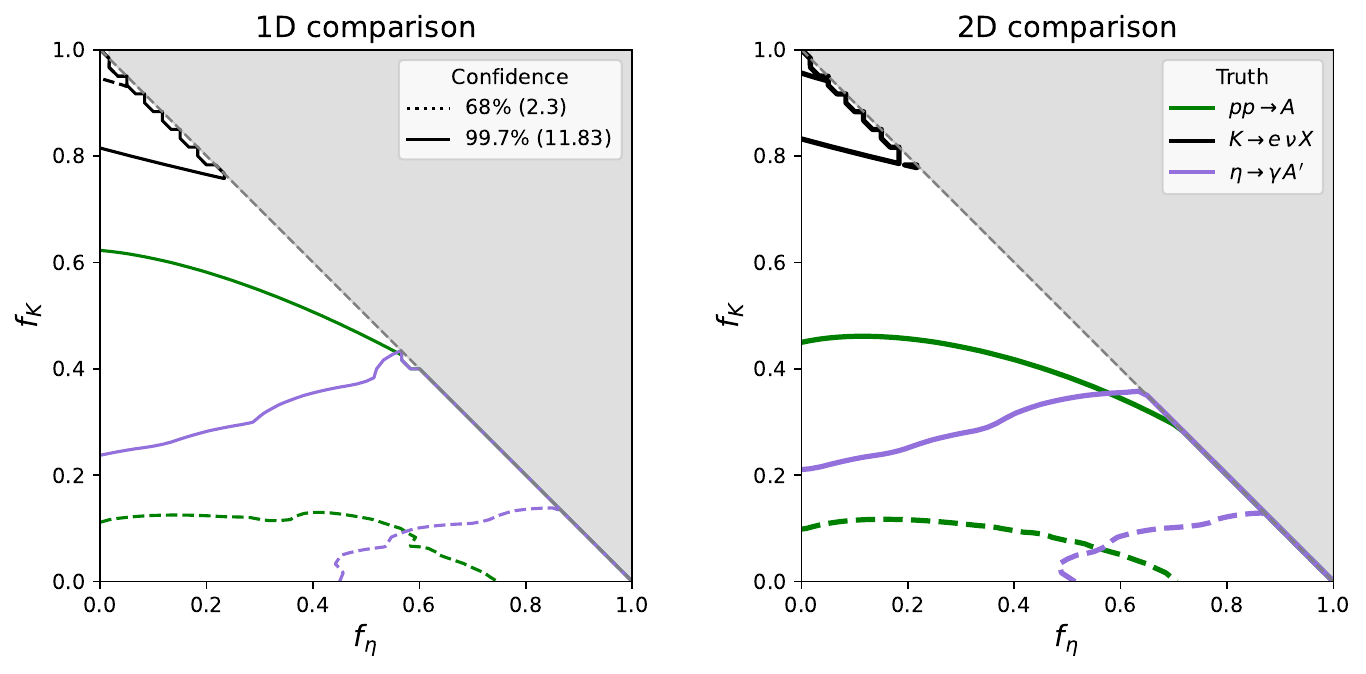}
\caption{Minimum likelihood comparison $\eta \rightarrow A'$ vs $pp \rightarrow A'$ vs $K\rightarrow \nu_e\, e\, Z'$ for $c\tau =10 m$\label{fig:chisq_3_comp_10m}}
\end{figure}

\begin{figure}[!htbp]
\centering
\includegraphics[width=\linewidth]{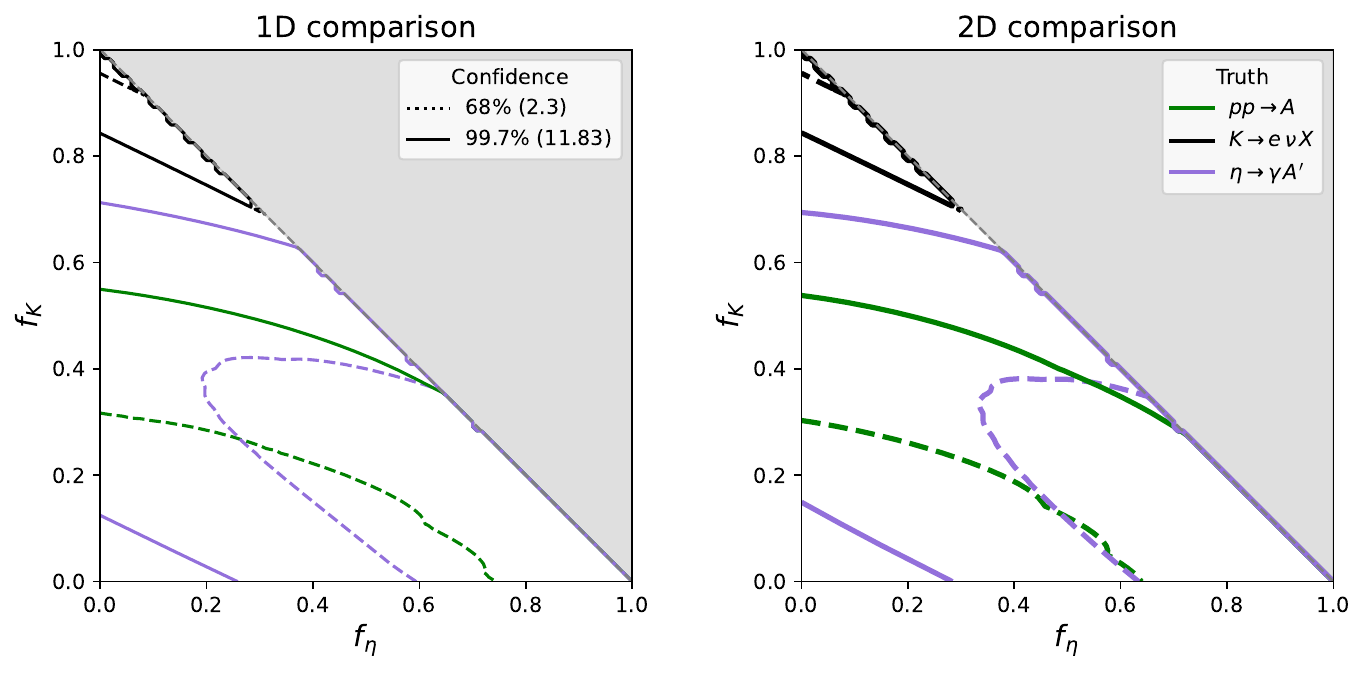}
\caption{Minimum likelihood comparison $\eta \rightarrow A'$ vs $pp \rightarrow A'$ vs $K\rightarrow \nu_e\, e\, Z'$ for $c\tau =100\, m$\label{fig:chisq_3_comp_100m}}
\end{figure}

We show further comparisons, assuming true LLP lifetimes of $\SI{10}{m}$ and $\SI{100}{m}$ in~\cref{fig:chisq_3_comp_10m} and~\cref{fig:chisq_3_comp_100m}, respectively.
Again, depending on the underlying truth, we find modest improvement when spatial information is incorporated in the 2D comparisons vs.~in the energy-only 1D comparisons.
%%%%
%%%%%%%%%%%%%%%%%%%%%%%%%%%%%%%%%%%%%%%%%
\bibliographystyle{utphys}
\bibliography{refs}

@article{Coloma:2023oxx,
    author = "Coloma, Pilar and Mart{\'\i}n-Albo, Justo and Urrea, Salvador",
    title = "{Discovering long-lived particles at DUNE}",
    eprint = "2309.06492",
    archivePrefix = "arXiv",
    primaryClass = "hep-ph",
    reportNumber = "IFT-UAM/CSIC-23-111, IFIC/23-40, FTUV-23-0823.4331",
    doi = "10.1103/PhysRevD.109.035013",
    journal = "Phys. Rev. D",
    volume = "109",
    number = "3",
    pages = "035013",
    year = "2024"
}

@article{Brdar:2025hqi,
    author = "Brdar, Vedran and Dutta, Bhaskar and Jang, Wooyoung and Kim, Doojin and Shoemaker, Ian M. and Tabrizi, Zahra and Thompson, Adrian and Yu, Jaehoon",
    title = "{Finding BSM Needles in Electromagnetic Haystacks at DUNE}",
    eprint = "2507.21228",
    archivePrefix = "arXiv",
    primaryClass = "hep-ph",
    reportNumber = "MI-HET-862, CETUP2025-007",
    month = "7",
    year = "2025"
}

@article{Batell:2023mdn,
    author = "Batell, Brian and Huang, Wenjie and Kelly, Kevin J.",
    title = "{Keeping it simple: simplified frameworks for long-lived particles at neutrino facilities}",
    eprint = "2304.11189",
    archivePrefix = "arXiv",
    primaryClass = "hep-ph",
    reportNumber = "CERN-TH-2023-030",
    doi = "10.1007/JHEP08(2023)092",
    journal = "JHEP",
    volume = "08",
    pages = "092",
    year = "2023"
}

@article{Bauer:2018onh,
    author = "Bauer, Martin and Foldenauer, Patrick and Jaeckel, Joerg",
    title = "{Hunting All the Hidden Photons}",
    eprint = "1803.05466",
    archivePrefix = "arXiv",
    primaryClass = "hep-ph",
    doi = "10.1007/JHEP07(2018)094",
    journal = "JHEP",
    volume = "07",
    pages = "094",
    year = "2018"
}

@article{Gori:2022vri,
    author = "Gori, Stefania and others",
    title = "{Dark Sector Physics at High-Intensity Experiments}",
    eprint = "2209.04671",
    archivePrefix = "arXiv",
    primaryClass = "hep-ph",
    reportNumber = "FERMILAB-PUB-22-672-SCD-T",
    month = "9",
    year = "2022"
}

@inproceedings{Batell:2022dpx,
    author = "Batell, Brian and Blinov, Nikita and Hearty, Christopher and McGehee, Robert",
    title = "{Exploring Dark Sector Portals with High Intensity Experiments}",
    booktitle = "{Snowmass 2021}",
    eprint = "2207.06905",
    archivePrefix = "arXiv",
    primaryClass = "hep-ph",
    month = "7",
    year = "2022"
}

@book{Gunion:1989we,
    author = "Gunion, John F. and Haber, Howard E. and Kane, Gordon L. and Dawson, Sally",
    title = "{The Higgs Hunter's Guide}",
    reportNumber = "SCIPP-89/13, UCD-89-4, BNL-41644",
    doi = "10.1201/9780429496448",
    isbn = "978-0-429-49644-8",
    volume = "80",
    year = "2000"
}

@article{Donoghue:1990xh,
    author = "Donoghue, John F. and Gasser, J. and Leutwyler, H.",
    title = "{The Decay of a Light Higgs Boson}",
    reportNumber = "CERN-TH-5644/90, BUTP-89-06",
    doi = "10.1016/0550-3213(90)90474-R",
    journal = "Nucl. Phys. B",
    volume = "343",
    pages = "341--368",
    year = "1990"
}

@article{Boiarska:2019jym,
    author = "Boiarska, Iryna and Bondarenko, Kyrylo and Boyarsky, Alexey and Gorkavenko, Volodymyr and Ovchynnikov, Maksym and Sokolenko, Anastasia",
    title = "{Phenomenology of GeV-scale scalar portal}",
    eprint = "1904.10447",
    archivePrefix = "arXiv",
    primaryClass = "hep-ph",
    doi = "10.1007/JHEP11(2019)162",
    journal = "JHEP",
    volume = "11",
    pages = "162",
    year = "2019"
}

@article{Winkler:2018qyg,
    author = "Winkler, Martin Wolfgang",
    title = "{Decay and detection of a light scalar boson mixing with the Higgs boson}",
    eprint = "1809.01876",
    archivePrefix = "arXiv",
    primaryClass = "hep-ph",
    reportNumber = "NORDITA-2018-087",
    doi = "10.1103/PhysRevD.99.015018",
    journal = "Phys. Rev. D",
    volume = "99",
    number = "1",
    pages = "015018",
    year = "2019"
}

@article{Bezrukov:2009yw,
    author = "Bezrukov, F. and Gorbunov, D.",
    title = "{Light inflaton Hunter's Guide}",
    eprint = "0912.0390",
    archivePrefix = "arXiv",
    primaryClass = "hep-ph",
    doi = "10.1007/JHEP05(2010)010",
    journal = "JHEP",
    volume = "05",
    pages = "010",
    year = "2010"
}

@article{Batell:2020vqn,
    author = "Batell, Brian and Evans, Jared A. and Gori, Stefania and Rai, Mudit",
    title = "{Dark Scalars and Heavy Neutral Leptons at DarkQuest}",
    eprint = "2008.08108",
    archivePrefix = "arXiv",
    primaryClass = "hep-ph",
    doi = "10.1007/JHEP05(2021)049",
    journal = "JHEP",
    volume = "05",
    pages = "049",
    year = "2021"
}

@article{Berlin:2018pwi,
    author = "Berlin, Asher and Gori, Stefania and Schuster, Philip and Toro, Natalia",
    title = "{Dark Sectors at the Fermilab SeaQuest Experiment}",
    eprint = "1804.00661",
    archivePrefix = "arXiv",
    primaryClass = "hep-ph",
    reportNumber = "SLAC-PUB-17238",
    doi = "10.1103/PhysRevD.98.035011",
    journal = "Phys. Rev. D",
    volume = "98",
    number = "3",
    pages = "035011",
    year = "2018"
}

@article{Okun:1982xi,
    author = "Okun, L. B.",
    title = "{LIMITS OF ELECTRODYNAMICS: PARAPHOTONS?}",
    reportNumber = "ITEP-48-1982",
    journal = "Sov. Phys. JETP",
    volume = "56",
    pages = "502",
    year = "1982"
}

@article{Holdom:1985ag,
    author = "Holdom, Bob",
    title = "{Two U(1)'s and Epsilon Charge Shifts}",
    reportNumber = "UTPT-85-30",
    doi = "10.1016/0370-2693(86)91377-8",
    journal = "Phys. Lett. B",
    volume = "166",
    pages = "196--198",
    year = "1986"
}

@article{deNiverville:2011it,
    author = "deNiverville, Patrick and Pospelov, Maxim and Ritz, Adam",
    title = "{Observing a light dark matter beam with neutrino experiments}",
    eprint = "1107.4580",
    archivePrefix = "arXiv",
    primaryClass = "hep-ph",
    doi = "10.1103/PhysRevD.84.075020",
    journal = "Phys. Rev. D",
    volume = "84",
    pages = "075020",
    year = "2011"
}

@article{Foot:1994vd,
    author = "Foot, Robert and He, X. G. and Lew, H. and Volkas, R. R.",
    title = "{Model for a light Z-prime boson}",
    eprint = "hep-ph/9401250",
    archivePrefix = "arXiv",
    reportNumber = "OITS-532, UM-P-93-115, OZ-93-26, IP-ASTP-32",
    doi = "10.1103/PhysRevD.50.4571",
    journal = "Phys. Rev. D",
    volume = "50",
    pages = "4571--4580",
    year = "1994"
}

@article{He:1990pn,
    author = "He, X. G. and Joshi, Girish C. and Lew, H. and Volkas, R. R.",
    title = "{NEW Z-prime PHENOMENOLOGY}",
    reportNumber = "UM-P-90/42, OZ-P-90/16",
    doi = "10.1103/PhysRevD.43.R22",
    journal = "Phys. Rev. D",
    volume = "43",
    pages = "22--24",
    year = "1991"
}

@article{Pospelov:2008zw,
    author = "Pospelov, Maxim",
    title = "{Secluded U(1) below the weak scale}",
    eprint = "0811.1030",
    archivePrefix = "arXiv",
    primaryClass = "hep-ph",
    doi = "10.1103/PhysRevD.80.095002",
    journal = "Phys. Rev. D",
    volume = "80",
    pages = "095002",
    year = "2009"
}

@online{Fields,
  author = {Laura Fields},
  title = {{DUNE} {F}luxes},
  year = 2021,
  howpublished = "{DUNE neutrino flux files generated with G4LBNF}",
  url = {https://glaucus.crc.nd.edu/DUNEFluxes/index.html},
}

@article{DUNE:2020ypp,
    author = "Abi, Babak and others",
    collaboration = "DUNE",
    title = "{Deep Underground Neutrino Experiment (DUNE), Far Detector Technical Design Report, Volume II: DUNE Physics}",
    eprint = "2002.03005",
    archivePrefix = "arXiv",
    primaryClass = "hep-ex",
    reportNumber = "FERMILAB-PUB-20-025-ND, FERMILAB-DESIGN-2020-02",
    month = "2",
    year = "2020"
}

@article{GEANT4:2002zbu,
    author = "Agostinelli, S. and others",
    collaboration = "GEANT4",
    title = "{GEANT4 - A Simulation Toolkit}",
    reportNumber = "SLAC-PUB-9350, FERMILAB-PUB-03-339, CERN-IT-2002-003",
    doi = "10.1016/S0168-9002(03)01368-8",
    journal = "Nucl. Instrum. Meth. A",
    volume = "506",
    pages = "250--303",
    year = "2003"
}

@article{Allison:2006ve,
    author = "Allison, John and others",
    title = "{Geant4 developments and applications}",
    reportNumber = "SLAC-PUB-11870",
    doi = "10.1109/TNS.2006.869826",
    journal = "IEEE Trans. Nucl. Sci.",
    volume = "53",
    pages = "270",
    year = "2006"
}

@article{Allison:2016lfl,
    author = "Allison, J. and others",
    title = "{Recent developments in Geant4}",
    reportNumber = "FERMILAB-PUB-16-447-CD",
    doi = "10.1016/j.nima.2016.06.125",
    journal = "Nucl. Instrum. Meth. A",
    volume = "835",
    pages = "186--225",
    year = "2016"
}

@article{Berryman:2019dme,
    author = "Berryman, Jeffrey M. and de Gouvea, Andre and Fox, Patrick J and Kayser, Boris Jules and Kelly, Kevin James and Raaf, Jennifer Lynne",
    title = "{Searches for Decays of New Particles in the DUNE Multi-Purpose Near Detector}",
    eprint = "1912.07622",
    archivePrefix = "arXiv",
    primaryClass = "hep-ph",
    reportNumber = "FERMILAB-PUB-19-607-ND-T, NUHEP-TH/19-16",
    doi = "10.1007/JHEP02(2020)174",
    journal = "JHEP",
    volume = "02",
    pages = "174",
    year = "2020"
}

@article{Dev:2021qjj,
    author = "Dev, P. S. Bhupal and Dutta, Bhaskar and Kelly, Kevin J. and Mohapatra, Rabindra N. and Zhang, Yongchao",
    title = "{Light, long-lived B {\ensuremath{-}} L gauge and Higgs bosons at the DUNE near detector}",
    eprint = "2104.07681",
    archivePrefix = "arXiv",
    primaryClass = "hep-ph",
    reportNumber = "FERMILAB-PUB-21-200-T, MI-TH-218",
    doi = "10.1007/JHEP07(2021)166",
    journal = "JHEP",
    volume = "07",
    pages = "166",
    year = "2021"
}

@article{Coloma:2020lgy,
    author = "Coloma, Pilar and Fern{\'a}ndez-Mart{\'\i}nez, Enrique and Gonz{\'a}lez-L{\'o}pez, Manuel and Hern{\'a}ndez-Garc{\'\i}a, Josu and Pavlovic, Zarko",
    title = "{GeV-scale neutrinos: interactions with mesons and DUNE sensitivity}",
    eprint = "2007.03701",
    archivePrefix = "arXiv",
    primaryClass = "hep-ph",
    reportNumber = "FERMILAB-PUB-20-269-ND",
    doi = "10.1140/epjc/s10052-021-08861-y",
    journal = "Eur. Phys. J. C",
    volume = "81",
    number = "1",
    pages = "78",
    year = "2021"
}

@article{Kelly:2020dda,
    author = "Kelly, Kevin J. and Kumar, Soubhik and Liu, Zhen",
    title = "{Heavy axion opportunities at the DUNE near detector}",
    eprint = "2011.05995",
    archivePrefix = "arXiv",
    primaryClass = "hep-ph",
    reportNumber = "FERMILAB-PUB-20-581-T",
    doi = "10.1103/PhysRevD.103.095002",
    journal = "Phys. Rev. D",
    volume = "103",
    number = "9",
    pages = "095002",
    year = "2021"
}

@article{Brdar:2020dpr,
    author = "Brdar, Vedran and Dutta, Bhaskar and Jang, Wooyoung and Kim, Doojin and Shoemaker, Ian M. and Tabrizi, Zahra and Thompson, Adrian and Yu, Jaehoon",
    title = "{Axionlike Particles at Future Neutrino Experiments: Closing the Cosmological Triangle}",
    eprint = "2011.07054",
    archivePrefix = "arXiv",
    primaryClass = "hep-ph",
    reportNumber = "FERMILAB-PUB-20-645-V, MI-TH-2029",
    doi = "10.1103/PhysRevLett.126.201801",
    journal = "Phys. Rev. Lett.",
    volume = "126",
    number = "20",
    pages = "201801",
    year = "2021"
}

@article{DUNE:2025lvs,
    author = "Abbaslu, S. and others",
    collaboration = "DUNE",
    title = "{Towards mono-energetic virtual $\nu$ beam cross-section measurements: A feasibility study of $\nu$-Ar interaction analysis with DUNE-PRISM}",
    eprint = "2509.07664",
    archivePrefix = "arXiv",
    primaryClass = "hep-ex",
    reportNumber = "FERMILAB-PUB-25-0627-LBNF",
    month = "9",
    year = "2025"
}

@article{Coloma:2019htx,
    author = "Coloma, Pilar and Hern{\'a}ndez, Pilar and Mu{\~n}oz, V{\'\i}ctor and Shoemaker, Ian M.",
    title = "{New constraints on Heavy Neutral Leptons from Super-Kamiokande data}",
    eprint = "1911.09129",
    archivePrefix = "arXiv",
    primaryClass = "hep-ph",
    doi = "10.1140/epjc/s10052-020-7795-z",
    journal = "Eur. Phys. J. C",
    volume = "80",
    number = "3",
    pages = "235",
    year = "2020"
}

@article{ICARUS:2024oqb,
    author = "Alrahman, F. Abd and others",
    collaboration = "ICARUS",
    title = "{Search for a Hidden Sector Scalar from Kaon Decay in the Dimuon Final State at ICARUS}",
    eprint = "2411.02727",
    archivePrefix = "arXiv",
    primaryClass = "hep-ex",
    reportNumber = "FERMILAB-PUB-24-0581-PPD",
    doi = "10.1103/PhysRevLett.134.151801",
    journal = "Phys. Rev. Lett.",
    volume = "134",
    number = "15",
    pages = "151801",
    year = "2025"
}

@article{BNL-E949:2009dza,
    author = "Artamonov, A. V. and others",
    collaboration = "BNL-E949",
    title = "{Study of the decay $K^+\to\pi^+\nu \bar\nu$ in the momentum region $140 < P_\pi < 199$ MeV/c}",
    eprint = "0903.0030",
    archivePrefix = "arXiv",
    primaryClass = "hep-ex",
    reportNumber = "BNL-81786-2008-JA, FERMILAB-PUB-09-007-CD-T, KEK-2008-44, TRIUMF-TRI-PP-08-26, UHEP-EX-08-004",
    doi = "10.1103/PhysRevD.79.092004",
    journal = "Phys. Rev. D",
    volume = "79",
    pages = "092004",
    year = "2009"
}

@article{NA62:2020pwi,
    author = "Cortina Gil, Eduardo and others",
    collaboration = "NA62",
    title = "{Search for $\pi^0$ decays to invisible particles}",
    eprint = "2010.07644",
    archivePrefix = "arXiv",
    primaryClass = "hep-ex",
    reportNumber = "CERN-EP-2020-193",
    doi = "10.1007/JHEP02(2021)201",
    journal = "JHEP",
    volume = "02",
    pages = "201",
    year = "2021"
}

@article{NA62:2021zjw,
    author = "Cortina Gil, Eduardo and others",
    collaboration = "NA62",
    title = "{Measurement of the very rare K$^{+}${\textrightarrow}$ {\pi}^{+}\nu \overline{\nu} $ decay}",
    eprint = "2103.15389",
    archivePrefix = "arXiv",
    primaryClass = "hep-ex",
    doi = "10.1007/JHEP06(2021)093",
    journal = "JHEP",
    volume = "06",
    pages = "093",
    year = "2021"
}

@article{Foroughi-Abari:2020gju,
    author = "Foroughi-Abari, Saeid and Ritz, Adam",
    title = "{LSND Constraints on the Higgs Portal}",
    eprint = "2004.14515",
    archivePrefix = "arXiv",
    primaryClass = "hep-ph",
    doi = "10.1103/PhysRevD.102.035015",
    journal = "Phys. Rev. D",
    volume = "102",
    number = "3",
    pages = "035015",
    year = "2020"
}

@article{Capozzi:2021nmp,
    author = "Capozzi, Francesco and Dutta, Bhaskar and Gurung, Gajendra and Jang, Wooyoung and Shoemaker, Ian M. and Thompson, Adrian and Yu, Jaehoon",
    title = "{Extending the reach of leptophilic boson searches at DUNE and MiniBooNE with bremsstrahlung and resonant production}",
    eprint = "2108.03262",
    archivePrefix = "arXiv",
    primaryClass = "hep-ph",
    reportNumber = "MI-HET-752",
    doi = "10.1103/PhysRevD.104.115010",
    journal = "Phys. Rev. D",
    volume = "104",
    number = "11",
    pages = "115010",
    year = "2021"
}

@article{MicroBooNE:2019izn,
    author = "Abratenko, P. and others",
    collaboration = "MicroBooNE",
    title = "{Search for Heavy Neutral Leptons Decaying into Muon-Pion Pairs in the MicroBooNE Detector}",
    eprint = "1911.10545",
    archivePrefix = "arXiv",
    primaryClass = "hep-ex",
    reportNumber = "FERMILAB-PUB-19-581-ND",
    doi = "10.1103/PhysRevD.101.052001",
    journal = "Phys. Rev. D",
    volume = "101",
    number = "5",
    pages = "052001",
    year = "2020"
}

@article{Shrock:1978ft,
    author = "Shrock, R. E.",
    title = "{A Test for the Existence of Effectively Stable Neutral Heavy Leptons}",
    reportNumber = "Print-78-0282 (PRINCETON)",
    doi = "10.1103/PhysRevLett.40.1688",
    journal = "Phys. Rev. Lett.",
    volume = "40",
    pages = "1688",
    year = "1978"
}

@article{FMMF:1994yvb,
    author = "Gallas, E. and others",
    collaboration = "FMMF",
    title = "{Search for neutral weakly interacting massive particles in the Fermilab Tevatron wide band neutrino beam}",
    doi = "10.1103/PhysRevD.52.6",
    journal = "Phys. Rev. D",
    volume = "52",
    pages = "6--14",
    year = "1995"
}

@article{T2K:2019jwa,
    author = "Abe, K. and others",
    collaboration = "T2K",
    title = "{Search for heavy neutrinos with the T2K near detector ND280}",
    eprint = "1902.07598",
    archivePrefix = "arXiv",
    primaryClass = "hep-ex",
    doi = "10.1103/PhysRevD.100.052006",
    journal = "Phys. Rev. D",
    volume = "100",
    number = "5",
    pages = "052006",
    year = "2019"
}

@article{NA482:2015wmo,
    author = "Batley, J. R. and others",
    collaboration = "NA48/2",
    title = "{Search for the dark photon in $\pi^0$ decays}",
    eprint = "1504.00607",
    archivePrefix = "arXiv",
    primaryClass = "hep-ex",
    reportNumber = "CERN-PH-EP-2015-093",
    doi = "10.1016/j.physletb.2015.04.068",
    journal = "Phys. Lett. B",
    volume = "746",
    pages = "178--185",
    year = "2015"
}

@article{LHCb:2019vmc,
    author = "Aaij, Roel and others",
    collaboration = "LHCb",
    title = "{Search for $A'\to\mu^+\mu^-$ Decays}",
    eprint = "1910.06926",
    archivePrefix = "arXiv",
    primaryClass = "hep-ex",
    reportNumber = "LHCb-PAPER-2019-031, CERN-EP-2019-212",
    doi = "10.1103/PhysRevLett.124.041801",
    journal = "Phys. Rev. Lett.",
    volume = "124",
    number = "4",
    pages = "041801",
    year = "2020"
}

@article{BaBar:2014zli,
    author = "Lees, J. P. and others",
    collaboration = "BaBar",
    title = "{Search for a Dark Photon in $e^+e^-$ Collisions at BaBar}",
    eprint = "1406.2980",
    archivePrefix = "arXiv",
    primaryClass = "hep-ex",
    reportNumber = "BABAR-PUB-14-002, SLAC-PUB-15979",
    doi = "10.1103/PhysRevLett.113.201801",
    journal = "Phys. Rev. Lett.",
    volume = "113",
    number = "20",
    pages = "201801",
    year = "2014"
}

@article{BaBar:2016sci,
    author = "Lees, J. P. and others",
    collaboration = "BaBar",
    title = "{Search for a muonic dark force at BABAR}",
    eprint = "1606.03501",
    archivePrefix = "arXiv",
    primaryClass = "hep-ex",
    reportNumber = "BABAR-PUB-16-003, SLAC-PUB-16549",
    doi = "10.1103/PhysRevD.94.011102",
    journal = "Phys. Rev. D",
    volume = "94",
    number = "1",
    pages = "011102",
    year = "2016"
}

@article{Tsai:2019buq,
    author = "Tsai, Yu-Dai and deNiverville, Patrick and Liu, Ming Xiong",
    title = "{Dark Photon and Muon $g-2$ Inspired Inelastic Dark Matter Models at the High-Energy Intensity Frontier}",
    eprint = "1908.07525",
    archivePrefix = "arXiv",
    primaryClass = "hep-ph",
    reportNumber = "FERMILAB-PUB-19-393-A-PPD",
    doi = "10.1103/PhysRevLett.126.181801",
    journal = "Phys. Rev. Lett.",
    volume = "126",
    number = "18",
    pages = "181801",
    year = "2021"
}

@article{Chang:2016ntp,
    author = "Chang, Jae Hyeok and Essig, Rouven and McDermott, Samuel D.",
    title = "{Revisiting Supernova 1987A Constraints on Dark Photons}",
    eprint = "1611.03864",
    archivePrefix = "arXiv",
    primaryClass = "hep-ph",
    reportNumber = "YITP-SB-16-44",
    doi = "10.1007/JHEP01(2017)107",
    journal = "JHEP",
    volume = "01",
    pages = "107",
    year = "2017"
}

@article{Foroughi-Abari:2021zbm,
    author = "Foroughi-Abari, Saeid and Ritz, Adam",
    title = "{Dark sector production via proton bremsstrahlung}",
    eprint = "2108.05900",
    archivePrefix = "arXiv",
    primaryClass = "hep-ph",
    doi = "10.1103/PhysRevD.105.095045",
    journal = "Phys. Rev. D",
    volume = "105",
    number = "9",
    pages = "095045",
    year = "2022"
}

@article{Gorbunov:2023jnx,
    author = "Gorbunov, Dmitry and Kriukova, Ekaterina",
    title = "{Dark photon production via elastic proton bremsstrahlung with non-zero momentum transfer}",
    eprint = "2306.15800",
    archivePrefix = "arXiv",
    primaryClass = "hep-ph",
    reportNumber = "INR-TH-2023-009",
    doi = "10.1007/JHEP01(2024)058",
    journal = "JHEP",
    volume = "01",
    pages = "058",
    year = "2024"
}
%%%%%%%%%%%%%%%%%%%%%%%%%%%%%%%%%%%%%%%%%

%%%%%%%%
\end{document}